\documentclass[prb,onecolumn,12pt]{revtex4-1}
\usepackage{epsfig}
\usepackage{amsmath}
\usepackage{amssymb}
\usepackage{threeparttable}
\usepackage{graphicx}
\usepackage{natbib}
\usepackage{longtable}
\usepackage{txfonts}
\usepackage{color}
\usepackage{braket}
\usepackage{revsymb}
\usepackage{float}
\usepackage{amsfonts,amssymb}
\usepackage{xr}
\usepackage{verbatim}
\begin{document}	

\title{High thermoelectric performances in PbP monolayers considering full electron-phonon coupling and four-phonon scattering processes}

\author{Ao Wu}
\affiliation{School of Information Science and Technology and Key Laboratory for Information Science of Electromagnetic Waves (MOE) and Department of Optical Science and Engineering and Key Laboratory of Micro and Nano Photonic Structures (MOE), Fudan University, Shanghai 200433, China}

\author{Yiming Zhang}	
\affiliation{School of Information Science and Technology and Key Laboratory for Information Science of Electromagnetic Waves (MOE) and Department of Optical Science and Engineering and Key Laboratory of Micro and Nano Photonic Structures (MOE), Fudan University, Shanghai 200433, China}

\author {Yujie Xia}
\affiliation{School of Information Science and Technology and Key Laboratory for Information Science of Electromagnetic Waves (MOE) and Department of Optical Science and Engineering and Key Laboratory of Micro and Nano Photonic Structures (MOE), Fudan University, Shanghai 200433, China}

\author {Lei Peng}
\affiliation{School of Information Science and Technology and Key Laboratory for Information Science of Electromagnetic Waves (MOE) and Department of Optical Science and Engineering and Key Laboratory of Micro and Nano Photonic Structures (MOE), Fudan University, Shanghai 200433, China}

\author{Hezhu Shao}\email{hzshao@wzu.edu.cn}
\affiliation{College of Electrical and Electronic Engineering, Wenzhou University, Wenzhou, 325035, China}

\author {Heyuan Zhu}
\affiliation{School of Information Science and Technology and Key Laboratory for Information Science of Electromagnetic Waves (MOE) and Department of Optical Science and Engineering and Key Laboratory of Micro and Nano Photonic Structures (MOE), Fudan University, Shanghai 200433, China}

\author{Hao Zhang} \email{zhangh@fudan.edu.cn}
\affiliation{School of Information Science and Technology and Key Laboratory for Information Science of Electromagnetic Waves (MOE) and Department of Optical Science and Engineering and Key Laboratory of Micro and Nano Photonic Structures (MOE), Fudan University, Shanghai 200433, China}
\affiliation{Yiwu Research Institute of Fudan University, Chengbei Road, Yiwu City, Zhejiang 322000, China}

\begin{abstract}
The band convergence strategy, which improves Seebeck coefficient by inducing multi-valley in bandstructures, has been widely used in thermoelectric performance (TE) enhancing. However, the phonon-assisted intervalley scattering effect is neglected and the mode-selection rules remain unclear. In this work, TE properties for $\alpha$-, $\beta$- and $\gamma$-PbP are intestigated under the consideration of full mode-, energy- and momentum-resolved electron-phonon interactions (EPI). The group theory is used to analyze the selection rules for EPI matrix elements. Our calculations reveal that, the intervalley scattering contributes non-trivially to the total carrier relaxation time, and the intervalley scattering can be modulated through crystal symmetry. In addition, the investigation on the thermal properties reveals that four-phonon scattering effect dominates the phonon relaxation processes, since the three-phonon scattering is suppressed due to the significantly large acoustic-optical phonon bandgap in $\alpha$-, $\beta$- and $\gamma$-PbP. By considering full EPI effect and high-order phonon scattering processes, the calculated ZT values reach 0.90, 0.24 and 1.25 for $\alpha$-, $\beta$- and $\gamma$-PbP, repectively, indicating their promising applications in thermoelectric devices.
\end{abstract}
	
\maketitle

\section{INTRODUCTION}
Thermoelectric materials, which can convert heat to electricity, has aroused extensive research interest due to the potential applications in clean and sustainable energy sources. The past decade witnesses great improvments in the thermoelectric (TE) utilities due to the discovering of many new high-performance TE materials as well as the propose of many effective strategies in TE enhancement like band convergence strategy. The TE merit $zT$ ($z T=S^{2} \sigma T /\left(\kappa_{\mathrm{e}}+\kappa_{\mathrm{l}}\right)$) is the principle parameter in quantifying the TE performance of a material, where $S$ is Seebeck coefficient, $\sigma$ is the electron conductivity, T is the absolute temperature and $\kappa_{\mathrm{e} / \mathrm{l}}$ is the electronic/lattice thermal conductivity. Generally,a good TE materials shall be a good electric conductor but poor thermal conductor. However, the parameters in the equation are coupled so tightly that it makes the optimization for $zT$ a challenging task.For example, large carrier concentration increases $\sigma$ and decrease $S$ simultaneously and the power factor, PF ($\mathrm{PF}=S^{2} \sigma \propto \mu m^{* 3 / 2}$) is hard to be regulated since the carrier mobility $\mu$ is inversely proportional to the density-of-state effective mass $m^*$. Nonetheless, many effort have been taken to contrive a compromise between the coupled quantities. Since the contribution from electron to thermal conductivity is neglectable especially in high-temperature zone, it is valid to pin hopes on materials with low lattice thermal conductivity. In addition, the $S$ ($S \propto m^{*}=N_{\mathrm{V}}^{2 / 3} m_{\mathrm{b}}^{*}$) can be further enhanced by increasing the band degeneracy, according to the band convergence strategy.

Single-layer PbP series of material is perdicted to be group IV and V compound with hexagonal structure. With the presentation of heavy atoms Pb, the phonon group velocity is decreased, indicating a low lattice thermal conductivity. Furthermore, the Debye temperature $\theta$ ($\theta={h v_{D}}/{k}$) is supposed to be low due to the low group velocity, which enables more phonons to active in high temperature and gives rise to large phonon scattering effects in high order phonon process and futher diminishes the lattice conductivity, thus making it a promising TE material candidates. For electronic part, PbP is reported to be narrow bandgap semiconductors, with bandgap below 1eV, where locates most high-performance TE materials. The reported bandstructure also demonstrate a so-called mexican-hat-shape dispersion at valance band maximum (VBM), giving rise to Van-Hove singularity in the density-of-state (DOS) near Fermi level.In this paper, we systematically study the electronic structures, thermo- and electro-dynamics and optical, transport and thermoelectric properties of the  lead phosphorene theoritically based on density functional theory (DFT) and density functional perturbation theory (DFPT).

\section{RESULTS AND DISCUSSION}

\subsection{Optimized crystal structure and structural stability}

\begin{figure}[ht!]
\centering
\includegraphics[width=1\linewidth]{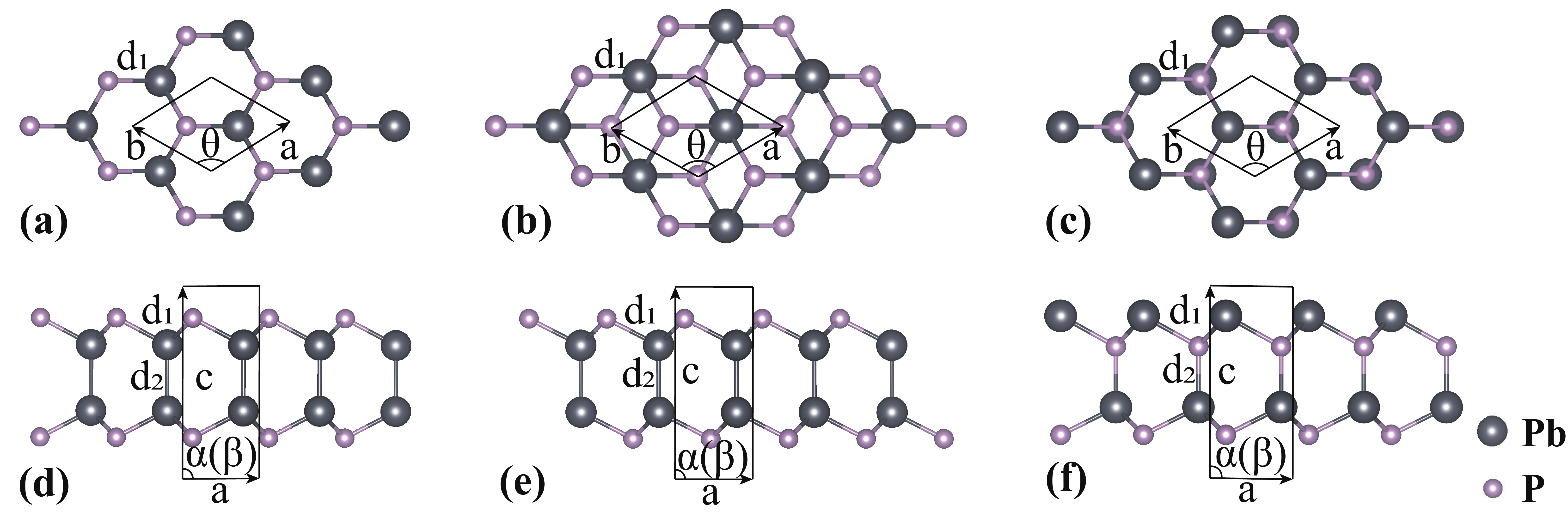}
\caption{Top view (a-c) and side view (d-f) of $\alpha$-PbP, $\beta$-PbP and $\gamma$-PbP}
\label{fig:structure}
\end{figure}

As shown in Figure~\ref{fig:structure}, monolayer $\alpha$-, $\beta$- and $\gamma$-PbP composed of two stacked sublayers belonging to hexagonal lattice, which are all low-buckled structures similar to those of silicene, germanene and blue phosphorene\cite{Matusalem2015,Liu2011,Liu2011a,Cahangirov2009,Cahangirov2009a}. $\alpha$-PbP with the space group of P-6m2 (\#187) possesses horizontal mirror symmetry and $\beta$-PbP with the space group of P-3m1 (\#164) possesses inversion symmetry. $\gamma$-PbP with the space group of P3m1 (\#156) is a janus structure by replacing the top or bottom sublayer with a different element in $\alpha$-PbP, which induces the break of the intrinsic out-of-plane symmetry, generally beneficial for some asymmetric properties such as Rashba-type spin splitting, second-harmonic generation, Zeeman-type spin splitting, and so on\cite{hou2020room}. The optimized lattice constants, bond lengths and bond angles are listed in Table~\ref{tab:structure}, and the optimized structural parameters of $\alpha$- and $\beta$-PbP monolayers are in good agreement with reported results\cite{Oezdamar2018}.

\begin{table}[ht!]
	\centering
	\caption{Structural and electronic properties of 2-D PbP in three phase. Lattice constant $a$, buckeling height $h_b$, bond length between Pb and P atoms in the same sublayer $d$, length between two sublayers $l$, bond angle $\theta$ and band gap energy calculated within PBE functions with spin-orbit coupling excluded $E_{PBE}$ and included $E_{PBE+SOC}$.}
	\renewcommand{\arraystretch}{1}
	\setlength\tabcolsep{8pt}
	\begin{tabular}{cccccccc}
		\hline
		\multicolumn{1}{c}{Structure} & \multicolumn{1}{c}{Space group} & \multicolumn{1}{c}{$a$(\r{A})} & \multicolumn{1}{c}{$d_1$(\r{A})} & \multicolumn{1}{c}{$d_2$(\r{A})} & \multicolumn{1}{c}{$\theta(deg)$} & \multicolumn{1}{c}{$Eg_{PBE}(eV)$} & \multicolumn{1}{c}{$Eg_{PBE+SOC}(eV)$}\\
		\hline
		$\alpha$-PbP  & P-6m2(\#187) & 4.09  & 2.66  & 3.06 & 100.16 & 0.50 & 0.51\\
		$\beta$-PbP   & P-3m1(\#164) & 4.13  & 2.68  & 3.08 & 100.88 & 0.39 & 0.41\\
		$\gamma$-PbP  & P3m1(\#156)  & 4.15  & 2.67  & 2.64 & 99.03  & 0.63 & 0.73\\
		\hline
	\end{tabular}%
	\label{tab:structure}%
\end{table}%

\begin{figure}[ht!]
\centering
\includegraphics[width=\linewidth]{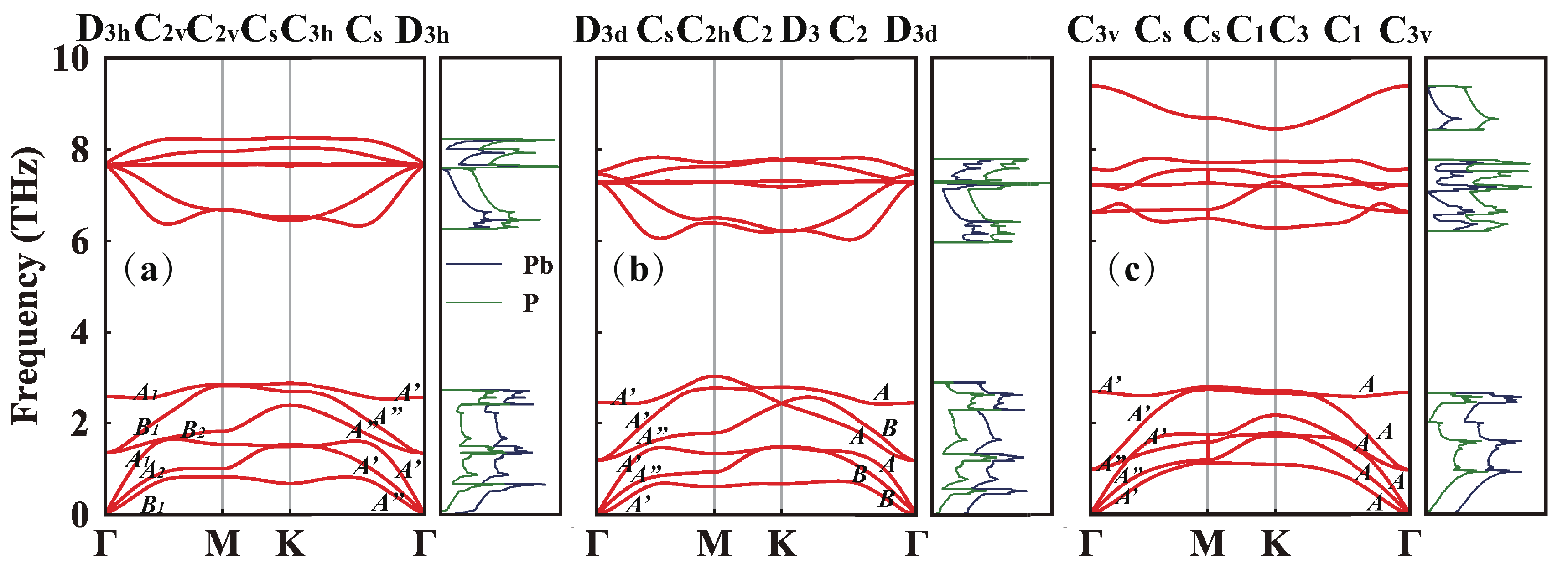}
\caption{(a-c) Phonon dispertion and partial density of state (PDOS) in log diagram at 300K in 4000ps of $\alpha$-PbP, $\beta$-PbP, $\gamma$-PbP respectively.}
\label{phonon} 
\end{figure}

\begin{figure}[ht!]
\centering
\includegraphics[width=1\linewidth]{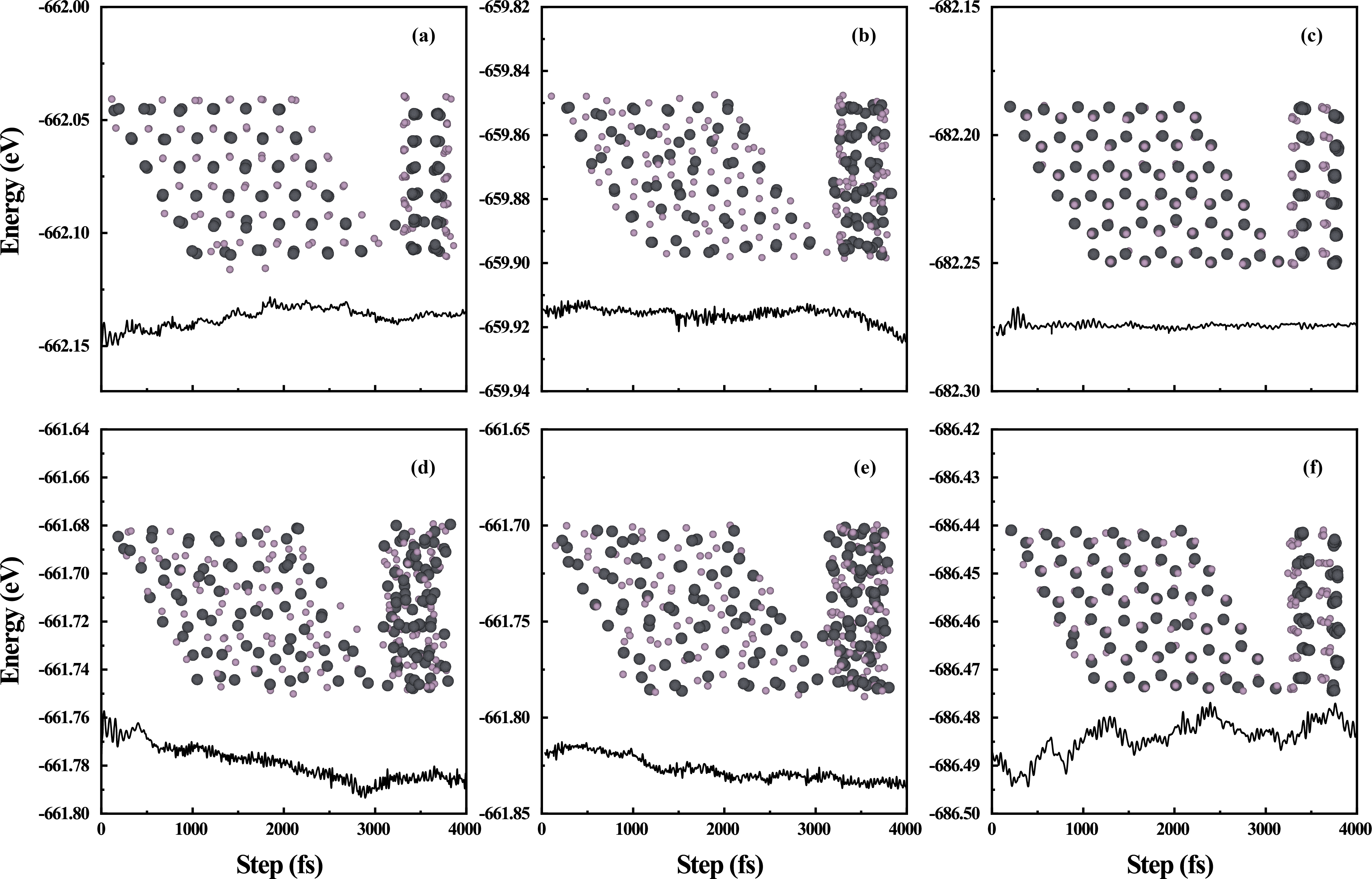}
\caption{The black plots represented the fluctuation of the total energy for $\alpha$-PbP, $\beta$-PbP and $\gamma$-PbP respectively at 300K (a-c) and 500K (d-f) within 4 $ps$, calculated by AIMD simulations. Snapshot of the terminal atomic configuration after the simulation is also provided.}
\label{md} 
\end{figure}

Figure~\ref{phonon}(a-c) shows the calculated phonon dispersions for the three monolayers, and no imaginary frequency can be observed, which indicates their thermal stabilities at low temperatures. We also perform the AIMD simulations on the three 2D PbP to investigate their thermodynamical stabilities at room temperature and the results are shown in Figure~\ref{md}(a-c), which reveals that, for the three monolayers, during the total simulation time of 4 ps with the time interval of 2 fs, the total energy fluctuates in a narrow range, and the atomic configurations only manifest a small deviation from their balanced positions, keeping well structural integrity. Therefore, the three 2D PhP maintain thermodynamical stabilities up to room temperature.

It should be noted that, as shown in Figure~\ref{phonon}(a-c), for the three monolayers, the three acoustic phonon branches and the three lower optical (quasi-acoustic) phonon branches are bunched together, and the rest optical phonon branches are bunched together, which finally generates a significantly large phonon bandgap seperating the quasi-acoustic and optical branches. The a-o phonon bandgaps for $\alpha$-, $\beta$- and $\gamma$-PbP monolayers are 3.43 THz, 2.93 THz and 3.43 THz, respectively. According to the calculated phonon partial density of states (PDOS) as shown in Figure~\ref{phonon}(a-c), the acoustic branches are mainly attributed from vibrations of heavier Pb element, and P atoms contribute dominantly to the rest optical branches, suggesting that, the large mass difference between Pb and P atoms induces the a-o phonon bandgap. Furthermore, since the phonon bandgap is larger than the maximum energy of the quasi-acoustic phonon branches, all the $aao$ three-phonon scattering channels are prohibited for these three monolayers restricted by the energy conservation\cite{Ravichandran2020}, leading to the significant suppressing of three-phonon scattering channels, and thus the enhancement of intrinsic lattice thermal conductivity $\kappa_L$. 

\subsection{Electronic bandstructures and chemical bonds}

\begin{figure}[ht!]
\centering
\includegraphics[width=0.7\linewidth]{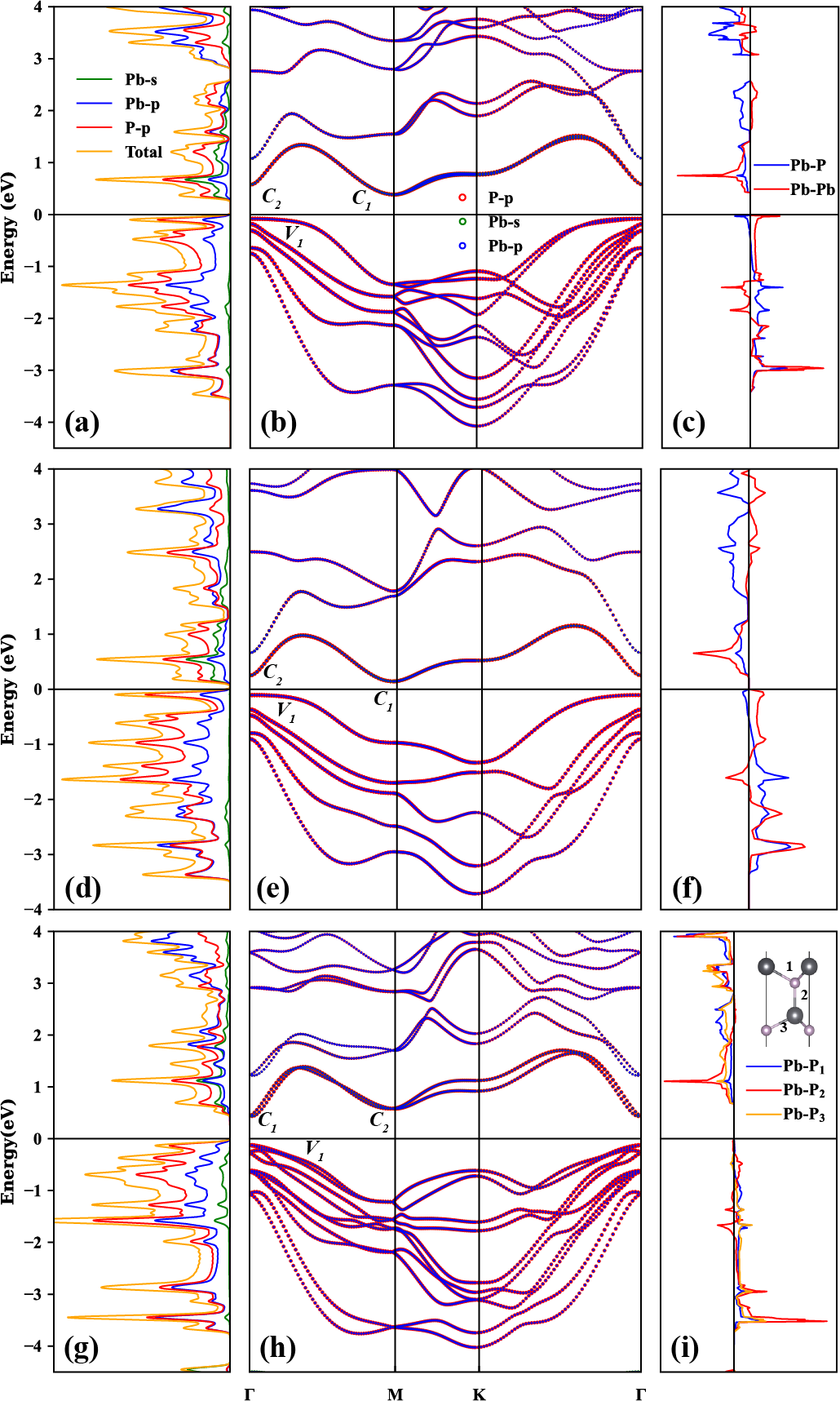}
\caption{(a-p) give the electronic band structures, partial density of states (PDOS) and crystal orbital hamitonian population (COHP) of $\alpha$-PbP, $\beta$-PbP, $\gamma$-PbP with the consideration of spin-orbit coupling (SOC). For gamma phase, Pb-P$_i$ (i=1,2,3) refer to three unequivalent Pb-P bonds as denoted in the sketch.}
\label{bandstructure} 
\end{figure}

\begin{figure}[ht!]
\centering
\includegraphics[width=1\linewidth]{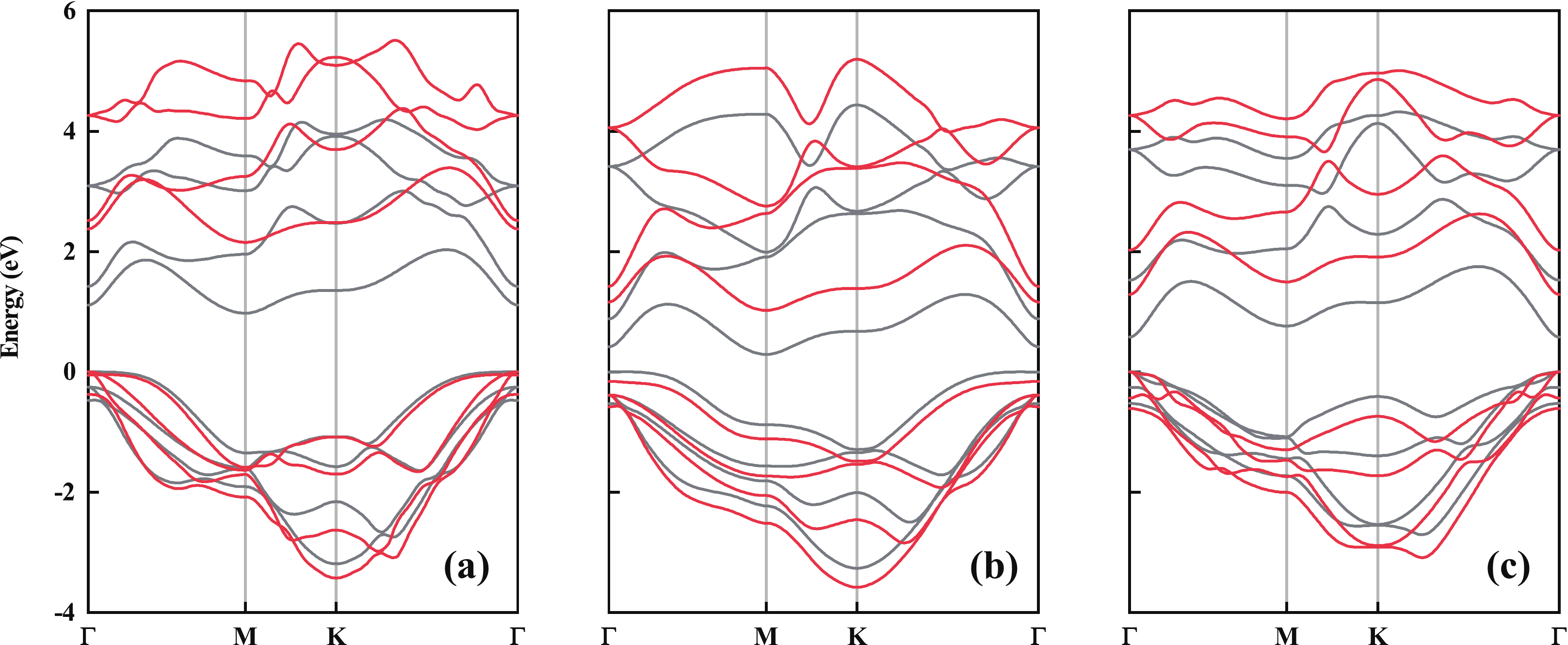}
\caption{Comparision of electronic band structures calculated with $G_0W_0$ (red) and LDA (grey).}
\label{gwband} 
\end{figure}

The calculated electronic band structures considering spin-orbit coupling (SOC) projected by atomic orbitals for these three PbP monolayers are shown in Figure~\ref{bandstructure}(a-c), which reveal that, $\alpha$-PbP and $\beta$-PbP are indirect-bandgap semiconductors with conduction-band minimum (CBM) located at $M$ point and valence-band maximum (VBM) located along $\Gamma-M$ high-symmetry line. $\gamma$-PbP is a direct-bandgap semiconductor with both CBM and VBM located at $\Gamma$ point. As a result of introducing the SOC due to the heavy Pb element, the spin degeneracy is lifted at $K$ point and other momemta for $\alpha$-PbP and $\gamma$-PbP. For non-magnetic $\beta$-PbP monolayer, which possesses time-reversal symmetry $\mathbb{R}$ ($\mathbb{R}:E_\uparrow(\mathbf{k})=E_\downarrow(\mathbf{-k})$) and inversion symmetry $\mathbb{I}$ ($\mathbb{I}:E_\uparrow(\mathbf{k})=E_\uparrow(\mathbf{-k})$) simultaneously, the spin degeneracy is protected by both symmetries $\mathbb{RI}$, as shown in Figure~\ref{bandstructure}(b). The calculated bandgaps based on the PBE+SOC method for these three PbP monolayers are 0.51 eV, 0.41 eV and 0.73 eV, respectively, as listed in Table~\ref{tab:structure}, which are increased to be 0.51 eV, 0.41 eV and 0.73 eV by considering the electronic many-body interactions based on the one-shot $G_0W_0$ method, as shown in Figure~\ref{gwband}. 

As shown in Figures~\ref{bandstructure}(b,e,h), for all the three monolayers, the highest valence bands are contributed mainly by the Pb-$p$ and P-$p$ orbitals, and the lowest conduction bands are mainly contributed from Pb-$s/p$ and P-$p$ orbitals, which are also verified by the partial density of state (PDOS) analysis as shown in Figures~\ref{bandstructure}(a,d,g). 
In order to further analyze the chemical bonds in these three PbP monolayers, we also conduct the -pCOHP (Projected Crystal Orbital Hamilton Populations) calculations for respective bonds implemented using the LOBSTER package\cite{Dronskowski1993,Deringer2011}, and the results are shown in Figures~\ref{bandstructure}(c,f,i), which indicate that, for $\alpha$-PbP and $\beta$-PbP, their antibonding and bonding states are contributed by the Pb-Pb bonds, whereas for $\gamma$-PbP, their antibonding and bonding states are composed mainly by the Pb-P bonds connecting the two sublayers. 

\subsection{Electronic transport: mode-resolved electron-phonon couplings and selection rules}

In semiconductors, the deformation potential approximation (DPA) method has been widely used to investigate the carrier transport properties for non-polar semiconductors, which only considers the interaction between carriers and longitudinal acoustic (LA) phonon in the long-wavelength limit. For the three PbP monolayers, $\alpha-$ and $\beta$-phases are non-polar and $\gamma$-phase is polar. The calculated carrier transport properties based on the DPA method are listed in Table~\ref{tab:DPAmobility}, which reveals that, the calculated intrinsic electron mobilities $\mu_e$ for $\alpha$- and $\beta$-PbP are large, i.e. $300\sim700~ \mathrm{cm^2/Vs}$, about one to two order of magnitude larger than the hole mobilities $\mu_h$ mainly due to the one order in magnitude larger of hole effective masses $m^*_h$ compared to electron effective masses $m^*_e$. For $\gamma$-PbP, due to its much larger deformation potential constant compared to $\alpha-$ and $\beta$-phases, the calculated carrier mobilities are much smaller.

\begin{table}
\centering
\caption{Calculated elastic modulus $\left(C^{2 D}\right)$, deformation potential constant $\left(D_{1}\right)$, effective mass $\left(m^{*}\right)$, and intrinsic carrier mobility $(\mu)$ in the $x$ (zigzag) and $y$ (armchair) directions of 2D-PbP at 300K}
\begin{tabular}{ p{2cm}p{2cm}p{2.4cm}p{1cm}p{1cm}p{1.4cm}p{1.4cm}p{2.5cm}p{2.5cm}  }
\hline
Materials & Directions & C$^{2D}$(N·m$^{-1}$) & D$_l$(e) & D$_l$(h) & m$^{*}_{e}$(m$_0$) & m$^{*}_{h}$(m$_0$) &$\mu$$_e$(cm$^2$V$^{-1}$s$^{-1}$) & $\mu$$_h$(cm$^2$V$^{-1}$s$^{-1}$)\\
\hline
$\alpha$-PbP   & x & 63.14 & 2.69 & 1.21 & 0.6310 & 12.4744& 311.58 & 3.97\\
& y & 63.50 & 3.70 & 1.18 & 0.3786 & 3.7129 & 460.08 & 4.74\\
$\beta$-PbP    & x & 58.40 & 2.34 & 1.11 & 0.4667 & 5.5581 & 696.23 & 21.82\\
& y & 56.53 & 3.15 & 1.06 & 0.4022 & 8.2767 & 500.72 & 10.54\\
$\gamma$-PbP   & x & 54.89 & 11.18 & 3.92 & 0.5634 & 0.7574 & 88.41 & 19.64\\
& y & 53.90 & 11.26 & 4.22 & 0.3480 & 0.7523 & 76.11 & 49.84\\
\hline
\end{tabular}
\label{tab:DPAmobility}
\end{table}


Since the DPA theory might misestimate the intrinsic carrier mobility especially for polar semiconductors or non-polar semiconductors in which the electron-phonon (\textit{el-ph}) couplings via other phonon modes dominate, further analysis regarding the full \textit{el-ph} couplings should be conducted to precisely describe the carrier transport properties in these materials. For $\alpha$-PbP monolayer with out-of-plane mirror reflection operation $\sigma_h$ similar to planar graphene ($D_{6h}$) or mirror-symmetric buckled structure MoS$_2$ ($D_{3h}$), the \textit{el-ph} scattering via odd ZA phonons are strongly suppressed, as the \textit{el-ph} elements involving odd ZA phonons are cancelled restrictedly by the mirror symmetry, according to the Mermin-Wagner theorem\cite{Nakamura2017,DSouza2020}. For polar $\gamma$-PbP, the Fr$\ddot{o}$hlich interaction between carriers and longitudinal optical (LO) phonon might play the key role in \textit{el-ph} interactions. 
In addition, as shown in Figure~\ref{bandstructure}, the bandstructures of the three PbP monolayers possess a single valley at $\Gamma$ point and three-fold-degeneracy valleys at $M$ point in the first Brillouin zone, which may induce non-trivial intervalley scatterings of electrons in these materials.

\begin{figure}[ht!]
\centering
\includegraphics[width=\linewidth]{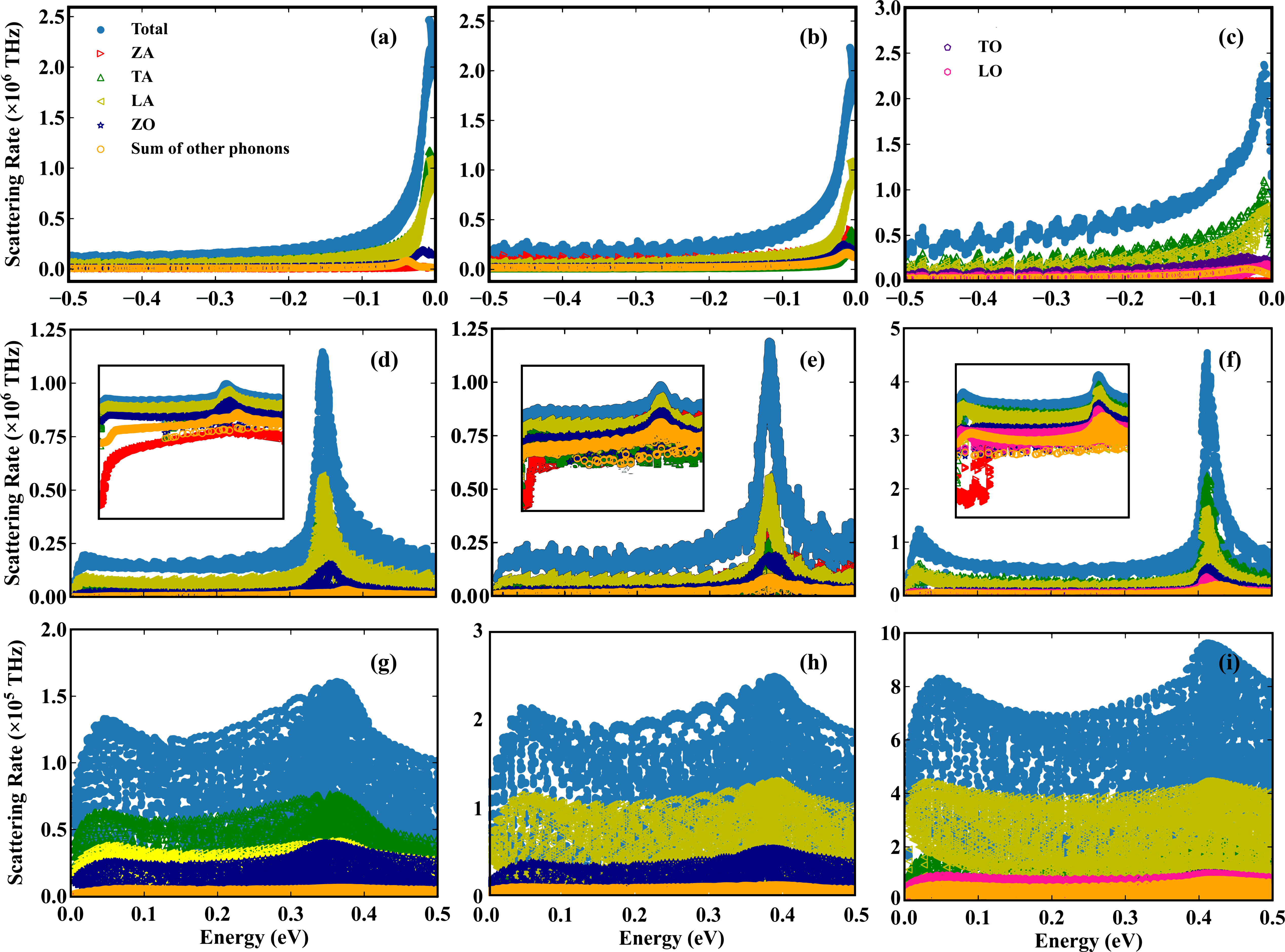}
\caption{Calculated electron-phonon scattering rate of (a-c) holes, (d-f) electrons (subplot in logarithmic scale) and (g-i) Seperated intravalley scattering rate of electrons for $\alpha$-PbP, $\beta$-PbP and  $\gamma$-PbP.}
\label{elphscatteringrate} 
\end{figure}

To precisely investigate the full \textit{el-ph} interactions in these three PbP monolayers, the mode-resolved \textit{el-ph} scattering rates $\left(1 / \tau_{n \mathbf{k}}\right)$ for electrons/holes with energy within 0.5 eV from CBM/VBM at 300 K are calculated, and the results are shown in Fig.~\ref{elphscatteringrate}, which reveals that, the total \textit{el-ph} scattering rates (denoted by solid blue dots) for these three monolayers are relatively large, approximately equal to ~2$\times10^{17}s^{-1}$, almost five orders larger than those in silicene and stanene\cite{Nakamura2017}, indicating strong \textit{el-ph} interactions in these materials. For $\alpha$-PbP monolayer possessing horizontal mirror symmetry as shown in Figure~\ref{elphscatteringrate}(a,d), LA and TA phonon modes dominate the \textit{el-ph} interactions, and ZA phonon modes are significantly surpressed due to the above-mentioned mirror-symmetry restriction. The peak of \textit{el-ph} scatterings around 0.35 eV from CBM is due to the large electron DOS as shown in Figure~\ref{bandstructure}(a). For $\beta$-PbP monolayer possessing inversion symmetry as shown in Figure~\ref{elphscatteringrate}(b,e), ZA/LA/TA phonon modes dominate the \textit{el-ph} interactions, and similar peak of \textit{el-ph} scatterings around 0.40 eV from CBM can be observed due to the enhanced electron DOS as well. For the polar janus $\gamma$-PbP monolayer as shown in Figure~\ref{elphscatteringrate}(c,f), TA, LA and TO$_1$ phonon modes dominate the \textit{el-ph} interactions, and the peak of \textit{el-ph} scatterings locates at 0.41 eV. In addition, the Fr$\ddot{o}$hlich interactions via LO$_1$ also contribute non-trivially.

\begin{figure}[ht!]
\centering
\includegraphics[width=\linewidth]{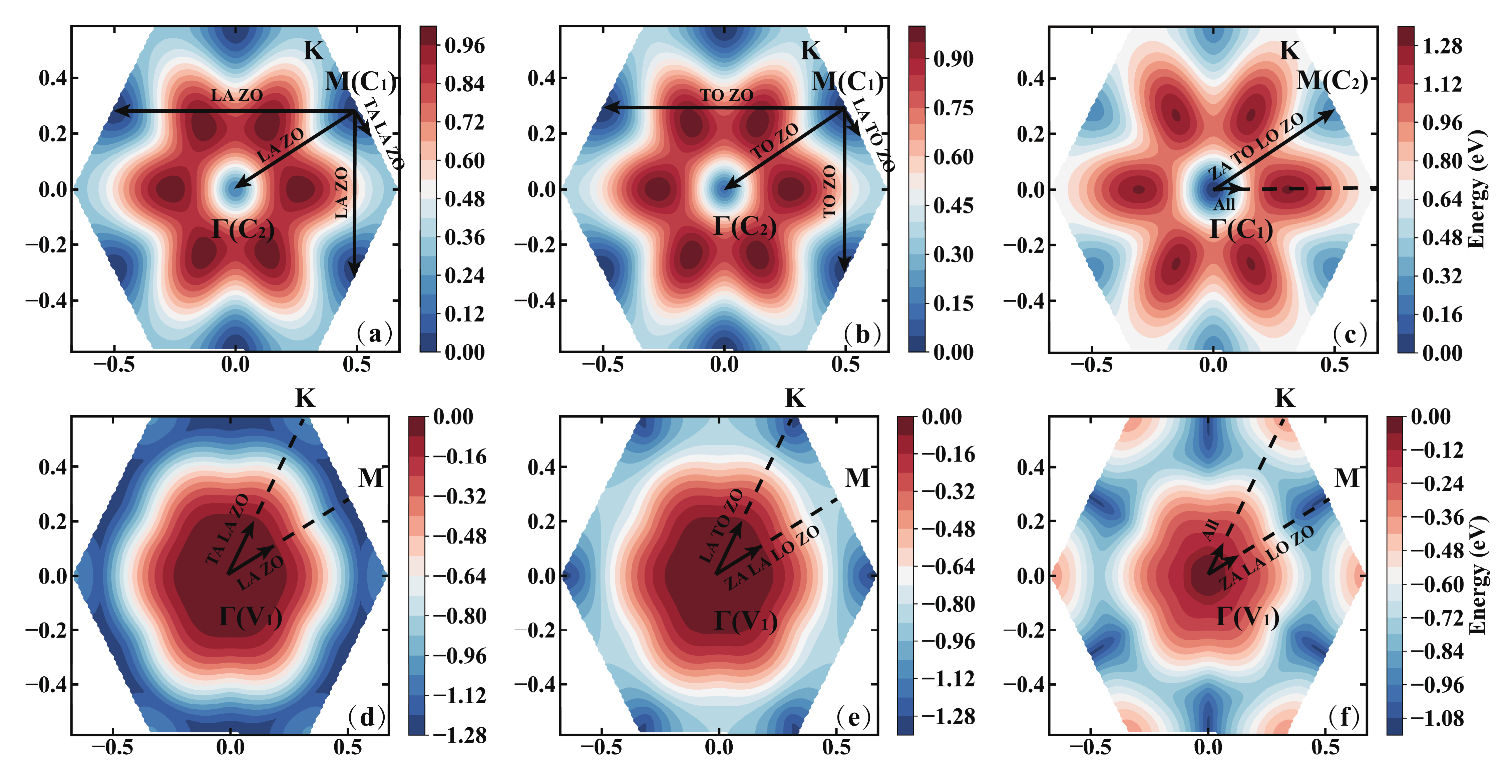}
\caption{Fermi surface contour map of the CB and VB of 2D-PbP in the first Brillouin zone in Cartesian coordinate and the allowed phonons for six lower phonon modes.}
\label{Contourmap} 
\end{figure}

In order to seperate the contribution from intravalley scattering and intervalley scattering, we further choose a small k-mesh region around the $M$ point and a dense q-mesh with small momenta around $\Gamma$ point\cite{Wu2021}, to describe the intravalley scatterings of CBM electrons for $\alpha$- and $\beta$-PbP monolayers. The calculated intravalley scatterings are shown in in Fig.~\ref{intravalleyscatteringrate}, which reveals that, the intravalley scatterings dominate the \textit{el-ph} interactions for electrons abround CBM. However, from electrons with energy of 0.32 eV from CBM, above the CBM, 
intervalley scattering begins to active, and finally reach a peak at ~0.35eV with the scattering rate overwhelming the intravalley scattering by about two times. Back to the shallow-dopping zone, where only intravalley scattering is active. The scattering rate increase almost linearly and after reaching a peak at ~0.17eV, it maintains stably at 2$\times10^{17}s^{-1}$ until 0.3eV. The electrons at the botton of $C_1$ can only be scattered by the phonons with highest energy, while with the Fermi level increases, phonons with lower energies, for example acoustic phonons, begin to meet the energy conservation criterion and scattering rate increase rapidly. When exceeded the phonon energy window, the el-ph scattering is totally influenced by energy and quasi-momentum conservation criterion and the particle DOS. As is for a constant temperature(300K), phonon population remains unchanged. However, electrons population varies with the doping concentration. From the band structure[Fig.~\ref{bandstructure}], DOS at valley $C_1$ and $C_2$ is quite low while a summit shows up at $K$ point, a saddle point in Brillouin zone. It is self-evident that the peak in DOS is responsible for the untra-large peak in the mode-resolved scattering rate at 0.35eV[Fig.~\ref{elphscatteringrate}]. With a large DOS, electrons can be scattered first to the meta-stable $K$ saddle point, and then to $C_1$ valley through relaxation process.

\begin{table}[ht!]
\centering
\caption{Selection rules for intravalley scattering of electrons in $\alpha-$/$\beta-$/$\gamma-$PbP.}
\renewcommand{\arraystretch}{1}
\setlength\tabcolsep{8pt}
\begin{tabular}{ccccccccc}
\hline
\multicolumn{1}{l}{Structure} & \multicolumn{1}{l}{Direction} &\multicolumn{1}{l}{Point group} & \multicolumn{1}{l}{$C_1^{'}$} & \multicolumn{1}{l}{$C_1^{''}$} & \multicolumn{1}{l}{Phonon irreps} & \multicolumn{1}{l}{Selected phonons}\\
\hline
$\alpha$-PbP & $M-\Gamma$ & ${C_{2v}}$ & ${B_2}$ & ${B_2}$  & ${A_1}$  &  ${LA}$ ${ZO_1}$ \\
             & $M-K$     & ${C_{s}}$  & ${A^{''}}$ & ${A^{''}}$  & ${A^{'}}$  & ${TA}$ ${LA}$ ${ZO_1}$\\
$\beta$-PbP  & $M-\Gamma$ & ${C_{s}}$ & ${A^{'}}$ & ${A^{'}}$  & ${A^{'}}$  &  ${ZA}$ ${LA}$ ${ZO_1}$ ${LO_1}$\\
             & $M-K$     & ${C_{2}}$ & ${B}$ & ${B}$  & ${A}$  &  ${TA}$ ${ZO_1}$ ${LO_1}$\\
$\gamma$-PbP & $M-\Gamma$ & ${C_{s}}$ & ${A^{'}}$ & ${A^{'}}$  & ${A^{'}}$  &  ${ZA}$ ${ZO_1}$ ${LO_1}$ ${TO_1}$\\
\hline
\end{tabular}%
\label{CBselectionrule}%
\end{table}%

\begin{table}[ht!]
\centering
\caption{Selection rules for intra-peak scattering of holes in $\alpha-$/$\beta-$/$\gamma-$PbP.}
\renewcommand{\arraystretch}{1}
\setlength\tabcolsep{8pt}
\begin{tabular}{cccccccccc}
\hline
\multicolumn{1}{l}{Structure} & \multicolumn{1}{l}{Direction} & \multicolumn{1}{l}{Point group} & \multicolumn{1}{l}{$V_1^{'}$} & \multicolumn{1}{l}{$V_1^{''}$} & \multicolumn{1}{l}{Phonon irreps} & \multicolumn{1}{l}{Selected phonons}\\
\hline
$\alpha$-PbP & $\Gamma-M$  & ${C_{2v}}$ & ${A_1}$    & ${A_1}$    & ${A_1}$    &  ${LA}$ ${ZO_1}$ \\
             & $\Gamma-K$  & ${C_{s}}$  & ${A^{'}}$  & ${A^{'}}$  & ${A^{'}}$  &  ${TA}$ ${LA}$ ${ZO_1}$\\
$\beta$-PbP  & $\Gamma-M$       & ${C_{s}}$  & ${A^{'}}$  & ${A^{'}}$  & ${A^{'}}$  &  ${ZA}$ ${LA}$ ${ZO_1}$ ${LO_1}$\\
             & $\Gamma-K$       & ${C_{2}}$  & ${A}$      & ${A}$      & ${A}$      &  ${TA}$ ${ZO_1}$ ${LO_1}$\\
$\gamma$-PbP & $\Gamma-M$       & ${C_{s}}$  & ${A^{''}}$ & ${A^{''}}$ & ${A^{'}}$  &  ${ZA}$ ${ZO_1}$ ${LO_1}$ ${TO_1}$\\
\hline
\end{tabular}%
\label{VBselectionrule}%
\end{table}%

To further reveal the underlying mechanisms for \textit{el-ph} interactions for these three monolayers, based on the perturbation-potential theory and group theory\cite{Malard2009}, we perform the analysis of selection rules for \textit{el-ph} interactions, which can be expressed as,

\begin{equation}
\mathrm{D}^{\mathrm{ph}} \otimes \mathrm{D}^{\mathrm{i}}=\sum_{v} \oplus a_{v} \mathrm{D}^{v}
\label{eq:selection}
\end{equation}

where $\mathrm{D}^{\mathrm{i}, \mathrm{ph}}$ represent the irreps of intial electron and phonon states. If the direct product of $\mathrm{D}^{\mathrm{i}}$ and $\mathrm{D}^{\mathrm{ph}}$ can be resolved into a linear decomposition including the final state $\mathrm{D}^{\mathrm{f}}$, or in other words, $a_{\mathrm{v}}\neq0$ , the transition is allowed. Otherwise, the transition is prohibited. 

For intra-peak scattering for holes in $\alpha$-PbP as shown in Fig.~\ref{bandstructure}(a) and Fig.~\ref{elphscatteringrate}(d), the point groups for states with momentum along $\Gamma-M$ and $M-K$ are $C_{2V}$ and $C_S$, respectively, sharing the same conjugate class $\sigma_h$ (out-of-plane mirror symmetry). Since there is no accidential band crossing in highest VB, the irreps of hole states around VBM are supposed to be the same or compatible in the whole Brillouin zone. Therefore, according to the selection rule as shown in Eq.~(\ref{eq:selection}), only phonon modes with identity representation (e.g. $A^{'}_{1}$ in $D_{3h}$, $A_1$ in $C_{2V}$ and $A^{'}$ in $C_s$ as listed in TABLE~\ref{excitonselectionrules}) can participate in the intravalley hole-scattering processes. In this way,  for the lower branches, LA/ZO$_1$ phonons are selected along $\Gamma$-$M$, and TA phonons are also selected along $\Gamma$-$K$ direction, which are in good agreement with the calculated results revealed in Fig.~\ref{elphscatteringrate}(d). ZA phonons are prohibited completely. For optic branches, ZO$_1$ phonons own significant contribution, but are less important compared to LA/TA branches. Similar analysis can be performed on the intra-peak scattering of holes in $\beta-$ and $\gamma-$PbP, and the results are listed in TABLE~S\ref{VBselectionrule}, which reveals that, 

Fig.~\ref{intravalleyscatteringrate} indicates that near the Fermi level, LA phonons can easily cater to the energy demands (steeper in phonon dispersion at long-wave approximation) and selection rules than TA phonons. The same analysis can be done to $\beta$-PbP and $\gamma$-PbP. For these two phases, ZA phonons is not prohibited. In addition, because of a lower symmetry, $\beta$-PbP and $\gamma$-PbP allows more phonon modes to interact with electrons. Since no band reversion is found in the three materials for CB and VB, the selection rules valid for CB and VB are the same. For VB, an even peak is found at $\Gamma$ point, with large electron DOS. The scattering rate reaches its peaks so quick that the calculation shows no ascent trend. Similarly, in-plane LA, TA phonons dominate the scattering for $\alpha$-PbP while flexual ZA is prohibited. 

Based on the calculated \textit{el-ph} interaction matrix elements and \textit{el-ph} scattering rates, the intrinsic temperature-dependent carrier mobilities can be calculated from the Boltzmann transport equation, which can be written as,

\begin{equation}
	\mu_{\alpha \beta}=\frac{-e}{n_{\mathrm{e}(\mathrm{h})} \Omega} \sum_{n \in \mathrm{CB}(\mathrm{VB})} \int \frac{\mathrm{d} \mathbf{k}}{\Omega_{\mathrm{BZ}}} \frac{\partial f_{n \mathbf{k}}^{0}}{\partial \varepsilon_{n} \mathbf{k}} v_{n \mathbf{k}, \alpha} v_{n \mathbf{k}, \beta} \tau_{n \mathbf{k}}
\end{equation}

\begin{figure}[ht!]
\centering
\includegraphics[width=\linewidth]{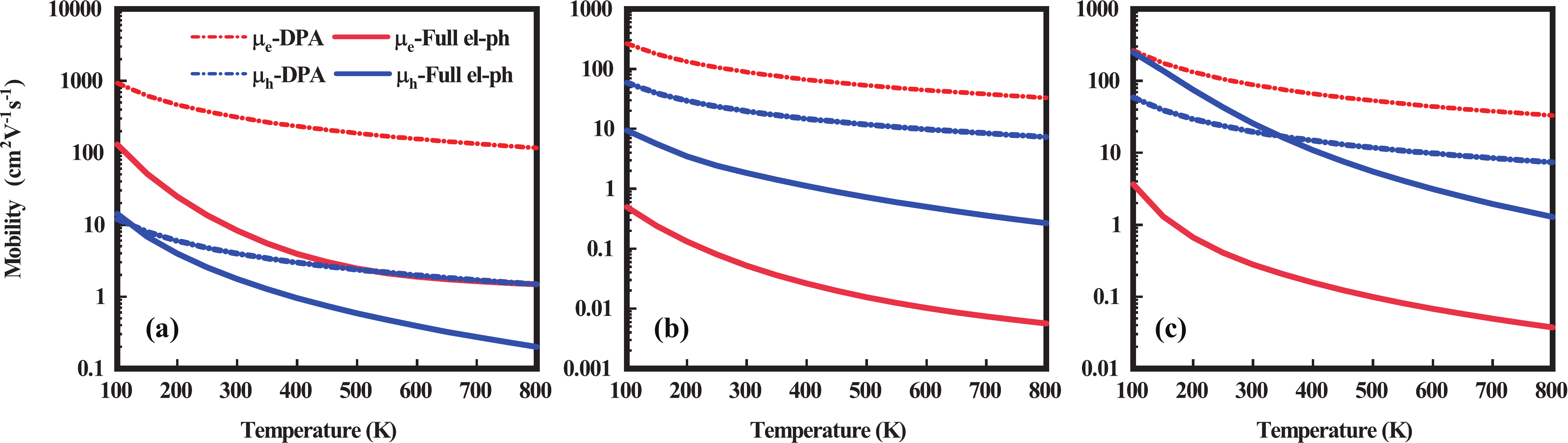}
\caption{Calculated carrier mobilities based on the DPA method and by considering full \textit{el-ph} scattering for (a) $\alpha$-, (b) $\beta$- and (c) $\gamma$-PbP.}
\label{mob} 
\end{figure}

where $n_{\mathrm{e}(\mathrm{h})}$ is the electron (hole) density, $\Omega$ and $\Omega_{\mathrm{BZ}}$ denote the volume of the unit cell and the first Brillouin zone, respectively.  $v_{\mathrm{nk}, \alpha}=h^{-1} \partial \varepsilon_{\mathrm{nk}} / \partial \mathrm{k}_{\alpha}$ denoting the velocity of the single-particle $\ket{n \mathbf{k}}$ electron along $\alpha$ direction. We carry out the calculations from 100 K to 800 K, and the results are shown in Fig.~\ref{mob} as solid lines. The mobility $\mu$ decreases as temperature increases, as more phonon modes are activated at higher temperatures. For comparision, the DPA-limited carrier mobilities are also calculated and denoted as dash lines. As mentioned above, the DPA method neglects the \textit{el-ph} scattering effects contributed from phonon modes other than LA phonons, including the optical branches that are activated at high temperatures, such as $ZO_1$ phonon modes in $\alpha$- and $\beta$-PbP and $TO_1$/$ZO_1$ phonon modes in $\gamma$-PbP, leading to the significant difference between mobilities calculated by the DPA method and those by considering full \textit{el-ph} scattering, especially at high temperatures. 
~Furthermore, as mentioned above, the DPA method also ignores the intervalley-scattering effects. However, as shown in Fig.~\ref{bandstructure} and  Fig~\ref{Contourmap}, although VBMs for all three monolayers and CBM for $\gamma-$PbP are located at $\Gamma$, at which only intravalley scattering effects are allowed due to the single degeneracy, the CBMs for $\alpha$- and $\beta$-PbP located at $M$ possess three-fold degeneracies, which allows intervalley scattering of electrons contributing to the total electron relaxation times, and thus makes the DPA results far from precision. This is consistent with those shown in Fig.~\ref{mob}, which reveals that, the two methods show larger deviation in electron mobilities $\mu_e$ compared to hole mobility $\mu_h$.


\begin{figure}[ht!]
\centering
\includegraphics[width=0.9\linewidth]{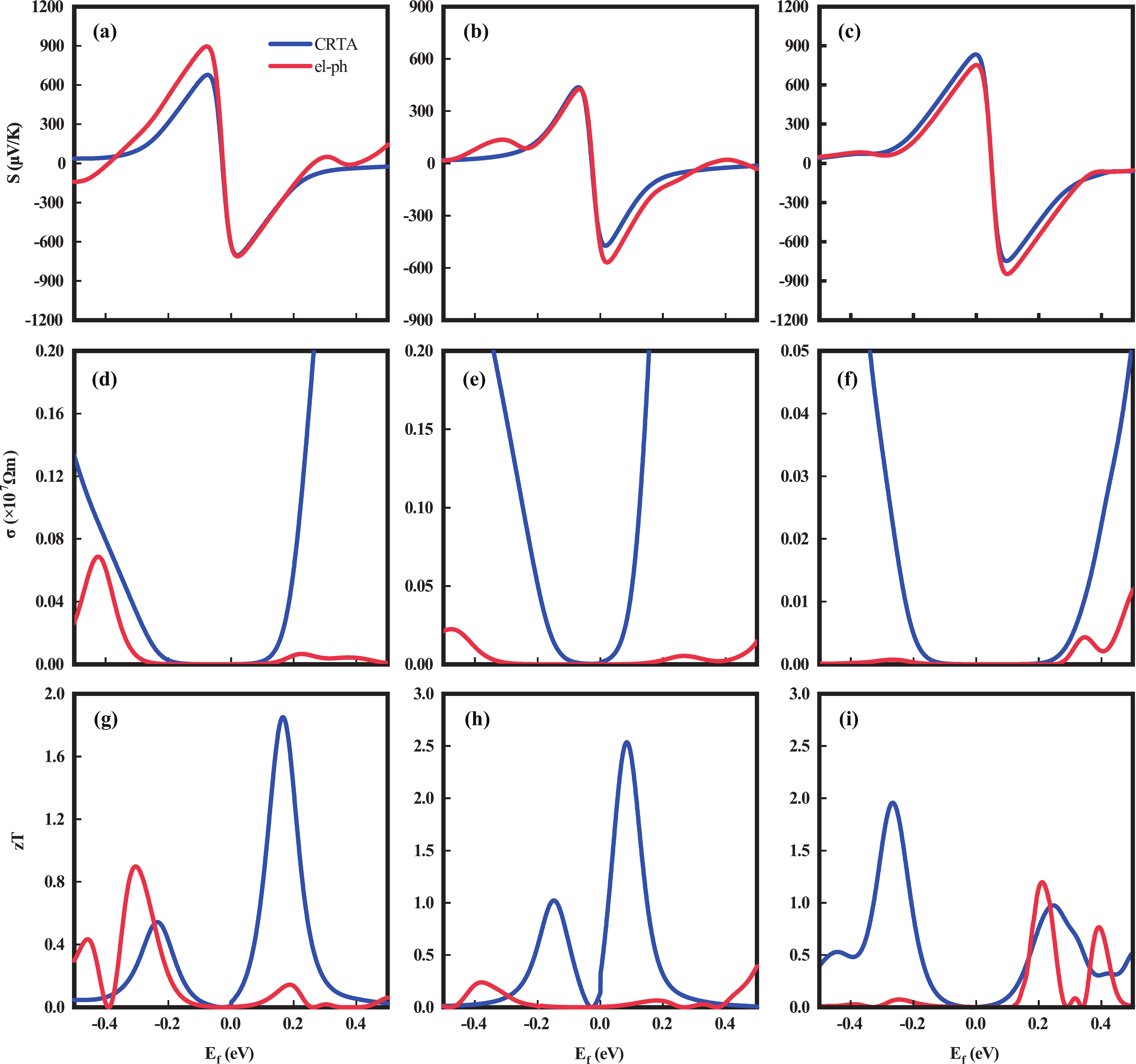}
\caption{Calculated Seebeck coefficient $S$ (a-c), electric conductivity $\sigma$ (d-f) and TE merit $zT$ (g-i) for $\alpha$-, $\beta$- and $\gamma$-PbP respectively}
\label{te} 
\end{figure}

The Seebeck coefficients S, electronic conductances $\sigma$ for these three monolayers calculated by solving the Boltzmann-transport equation based on the rigid-band approximation implemented in the BoltzTraP2 according to Eqs.~(\ref{eq:nh}-\ref{eq:seebeck}) are shown in Fig~\ref{te}(a-c) and (d-f), and to deal with the mode-resolved carrier relaxation time $\tau_{n\mathbf{k}}$, both the constant relaxation time approximation (CRTA) and full \textit{el-ph} interactions are used for comparision. The constant electronic relaxation time $\tau_0$ used in the CRTA method is obtained according to the DPA method as listed in Table ~\ref{tab:DPAmobility}. The calculated Seebeck coefficients reach nearly 900 $\mu V/K$ for $\alpha$- and $\gamma$-PbP and 450 $\mu V/K$ for $\beta$-PbP close to the CBM and VBM, which are much higher than conventional TE materials such as PbTe (185 $\mu V/K$), Bi$_2$Te$_3$ (215 $\mu V/K$) and SnSe ($\sim$510 $\mu V/K$). 
The S can be described by the Mott relationship\cite{sun2015},

\begin{equation}
S=-\frac{\pi^{2}}{3} \frac{k_{\mathrm{B}}^{2} T}{e}\left[\frac{\partial \ln N(E)}{\partial E}+\frac{\partial \ln \tau(E)}{\partial E}\right]_{E_{\mathrm{f}}}
\label{eq:seebeck}
\end{equation}

where $N(E)$ and $\tau(E)$ are the energy dependent DOS and electronic relaxation time. The Seebeck coefficients S is comparable for electrons and holes even though the valence band is much flatter than conduction band (larger energy dependent DOS) due to a high CBM valley degeneracies. Also, the Seebeck coefficients is insensitive to electron-phonon couplings as the relaxation time remains basically unchanged as the energy shift. 

However, the electronic conductances depends greatly on the \textit{el-ph} couplings as are closely related to carrier mobilities via $\sigma=n e \mu$, where $n$ is the carrier concentration.

The total thermal conductivities $\kappa$ are generally composed of lattice thermal conductivities $\kappa_l$ and electronic thermal conductivities $\kappa_e$.  The electronic thermal conductivities $\kappa_e$ can be calculated based on the Wiedermann-Franz law, i.e. $\kappa_e =L\sigma T$, where $L$ is the Lorenz constant. At low temperatures, phonons are dominantly scattered by impurities and boundaries, where $\kappa_{l} \propto T^{2}$ in 2D systems according to the Debye model. When the temperature is much higher than Debye temperature $\Theta$, all the phonons are activated and the Umklapp processes of phonon-phonon scatterings dominate, thus $\kappa_{1} \propto  1 / T$\cite{Block2021}. The presence of Pb atoms significantly lowers the frequencies of acoustic phonon modes and gives rise to a Debye temperature $\Theta$ well below room temperature, i.e. 134.38 K, 143.98 K and 139.18 K for $\alpha$-, $\beta$- and $\gamma$-PbP calculated according to Eq.~\ref{}. Therefore, herein we only take in consideration the intrinsic lattice thermal conductivities with the correction of four-phonon interaction processes in thermal conductivities.

\subsection{Thermal transport and thermoelectric performance: four-phonon interactions}

Generally, materials with good electronic transport properties and low thermal conductivities simultaneously are good candidates for thermoelectric applications. In semiconductors, the lattice thermal conductivity $\kappa_L$ can be calculated based on the kinetic theory\cite{Lindsay2013,Lindsay2014},

\begin{equation}
\kappa_{\alpha \beta}=\frac{1}{V} \sum_{\lambda} C_{\lambda} v_{\lambda \alpha} v_{\lambda \beta} \tau_{\lambda}
\label{eq:kappal}
\end{equation}

where $V$ is the crystal volume, $\lambda$ is the phonon mode with both wave vector $\textbf{q}$ and phonon branch index $\nu$. i.e. $\lambda=(\mathbf{q}\nu)$. $C_\lambda$ is the mode-resolved heat capacity, $\nu _{\lambda \alpha}$ ($\nu _{\lambda \beta}$) is the group velocity of phonon mode $\ket{\lambda}$ along $\alpha$ ($\beta$) direction, which is given by $v_{\lambda \alpha}=\frac{d \omega_{\lambda}}{d q_{\alpha}}$, with $\omega_{\lambda}$ representing the phonon frequency. Phonon scattering generally includes isotopes scattering ($\tau_{iso}$), boundary scattering ($\tau_B$), phonon-phonon scattering such as three-phonon scattering ($\tau_{3ph}$), four-phonon scattering ($\tau_{4ph}$) in some materials possessing significant a-o phonon bandgap, and etc, which together contribute to the total mode-resolved phonon relaxation time $\tau_\lambda$ via the Matthiessen's rule,

\begin{equation}
\frac{1}{\tau_\lambda}=\frac{1}{\tau_{\lambda,iso}}+\frac{1}{\tau_{\lambda.B}}+\frac{1}{\tau_{\lambda,3ph}}+\frac{1}{\tau_{\lambda,4ph}}
\label{eq:tau}
\end{equation}

\begin{figure}[ht!]
\centering
\includegraphics[width=\linewidth]{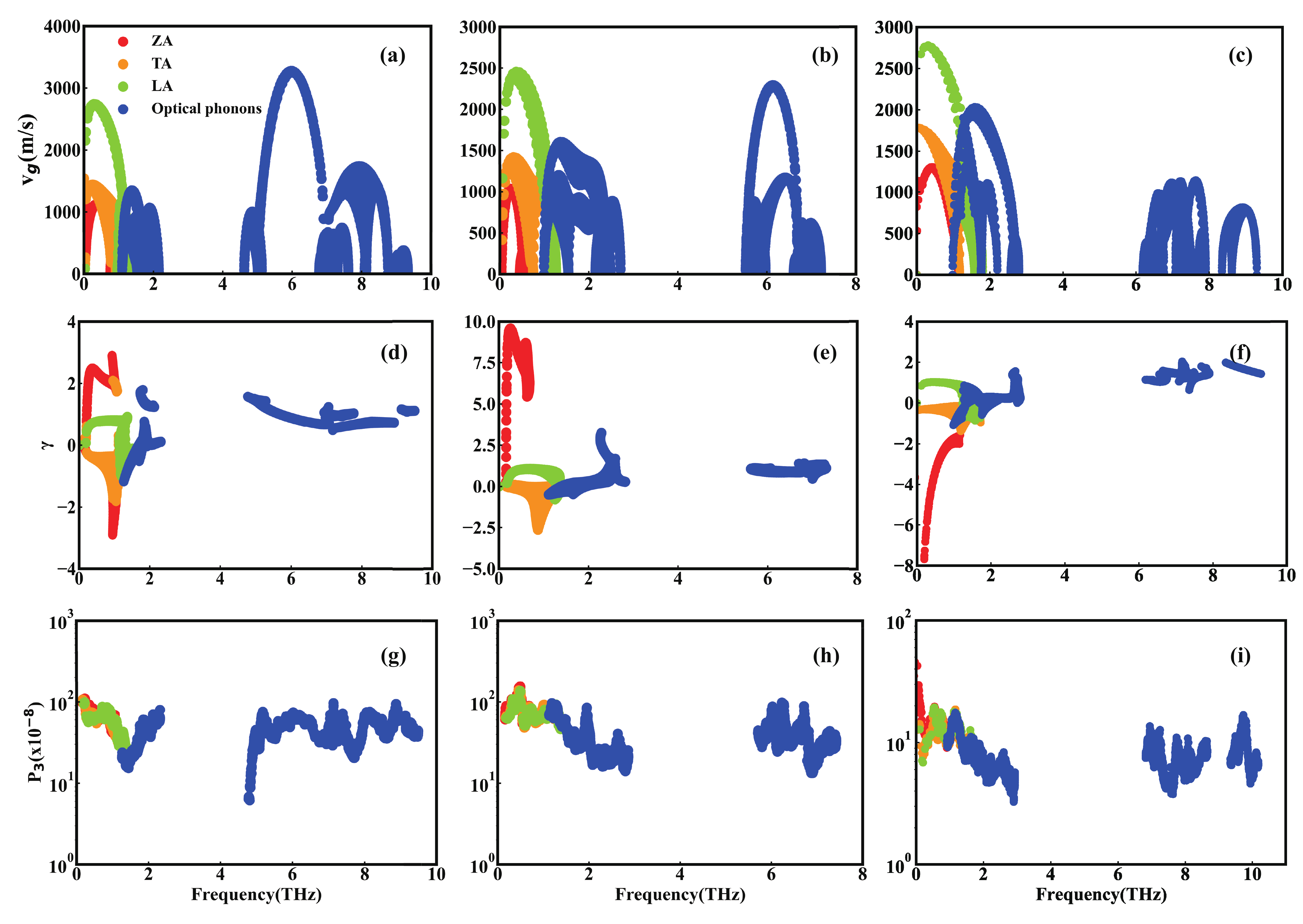}
\caption{Calculated thermal properties of the three structures $\alpha$-PbP, $\beta$-PbP and $\gamma$-PbP: (a-c) Phonon group velocities, (d-f) mode Gruneisen parameters and (g-i) three phonons scattering phase space.}
\label{fig:thermal} 
\end{figure}

\begin{figure}[ht!]
\centering
\includegraphics[width=\linewidth]{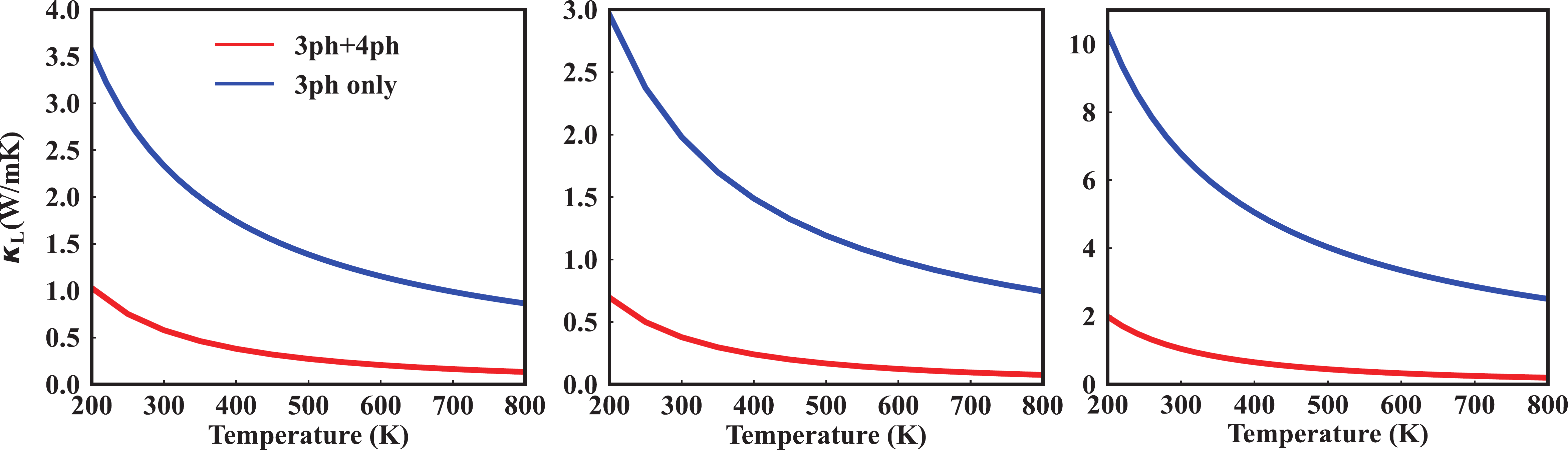}
\caption{(a-c) Calculated intrinsic lattice thermal conductivity for $\alpha$-PbP, $\beta$-PbP and $\gamma$-PbP in consideration of three-phonon interactions only and both three- and four-phonon interactions.}
\label{fig:kappa} 
\end{figure}

Generally, the scattering rates of phonon are determined by two factors\cite{Wu2016}: (i) the strength of anharmonic phonon-phonon interactions measured by Gruneisen parameters, and (ii) the number of phonon-phonon scattering channels\cite{Li2014,Lindsay2008,Peng2016}. The calculated group velocities $v_g$, Gruneisen parameters $\gamma$ and three-phonon phase space $P_3$ for $\alpha-$, $\beta-$ and $\gamma-$PbP are shown in Figure~\ref{fig:thermal}, which reveals that, the group velocities $v_g$ of $\alpha-$PbP are generally larger than those of $\beta-$PbP, which are larger than those of $\gamma-$PbP, and the Gruneisen parameters $\gamma$ indicating the strength of anharmonic interactions of $\gamma-$PbP are stronger than those of $\beta-$PbP, which are stronger than those of $\alpha-$PbP, and finally, the three-phonon phase spaces $P_3$ of $\alpha-$PbP are comparable to those of $\beta-$PbP, which are larger than those of $\gamma-$PbP roughly by an order in magnitude as shown in Figure~\ref{fig:thermal}(g-i). Then, the three-phonon-limited lattice thermal conductivities $\kappa_l$ can be calculated and the results are shown in Figure~\ref{fig:kappa} denoted as the blue lines, which reveal that, the calculated $\kappa_L$ at room temperature are 2.34 $W/mK$, 1.98 $W/mK$ and 6.78 $W/mK$ for $\alpha-$, $\beta-$ and $\gamma-$PbP respectively. The $\kappa_L$ for $\gamma-$PbP is larger than those of $\alpha-$ and $\beta-$PbP by several times, which is probably due to its much larger three-phonon phase space $P_3$.

As mentioned above, due to the mass difference between Pb and P atoms and the bunching effects of quasi-acoustic phonon branches, significantly large a-o phonon bandgaps can be observed in all three monolayers as shown in Figure~\ref{phonon}, and the maximum frequency of quasi-acoustic phonon modes is smaller than the a-o phonon bandgap, leading to the suppression of $aao$ scattering channels in three-phonon scattering processes. Therefore, similar to the case of boron arsenide and following the proposed new criteria about high $\kappa_L$ for binary semiconductors, i.e. (i) large a-o gap mainly due to large mass difference between constituent atoms, (ii) bunching effects of acoustic phonon branches and (iii)isotopically pure heave atoms, the $\kappa_L$ for $\alpha-$, $\beta-$ and $\gamma-$PbP are believed to be significantly influenced by four-phonon scattering processes, which obey the energy conservation and are thus allowed.


\begin{figure}[ht!]
\centering
\includegraphics[width=0.7\linewidth]{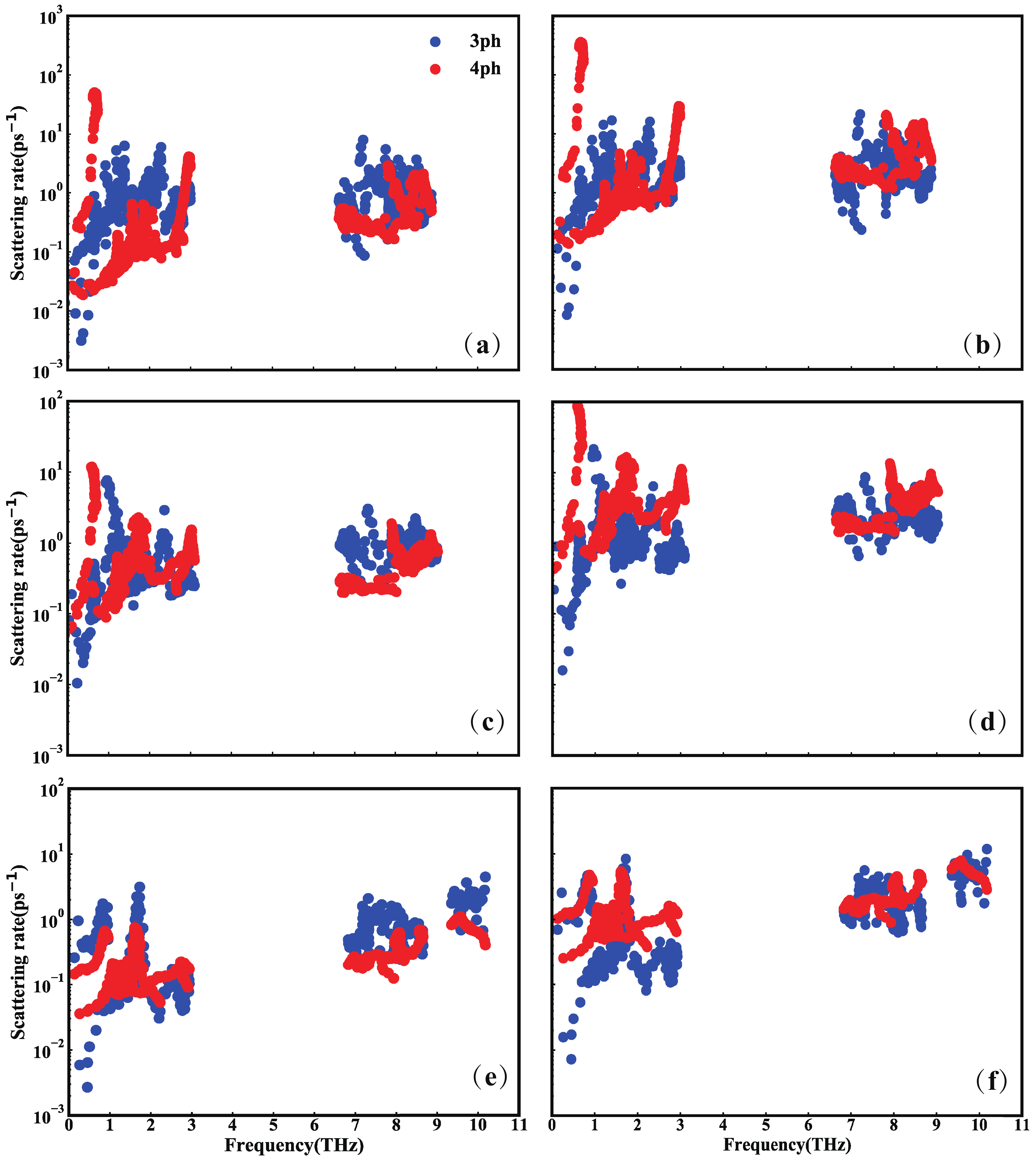}
\caption{Calculated three-phonon and four-phonon process scattering rate for (a,b) $\alpha$-PbP, (c,d) $\beta$-PbP and (e,f) $\gamma$-PbP at 300 K and 800 K.}
\label{fig:phscattering} 
\end{figure}


For comparison, the three- and four-phonon scattering rates at 300 K and 800 K for all three monolayers are calculated and the results are shown in Figure~\ref{fig:phscattering}, which reveals that, at 300 K, the four-phonon scattering rates of quasi-ascoutic phonon modes in $\alpha-$PbP are roughly smaller than the corresponding three-phonon scattering rates, but they are comparable in $\beta-$ and $\gamma-$PbP. The four-phonon scattering rates of optical phonon modes in $\gamma-$PbP are roughly smaller than the corresponding three-phonon scattering rates, but they are comparable in $\alpha-$ and $\beta-$PbP. When the temperature increases to 800 K, both three- and four-phonon scattering rates increase as well, but seemingly four-phonon scattering rates increase further. At 800 K, for quasi-acousti phonon modes in $\beta-$ and $\gamma-$PbP, four-phonon scattering rates surpass three-phonon scattering rates, and they are nearly comparable in $\alpha-$PbP. For optical phonon modes in all three monolayers, four-phonon scattering rates are comparable to three-phonon scattering rates. The calculated $\kappa_L$ for all three monolayers by considering both three- and four-phonon scattering processes are shown in Figure~\ref{fig:kappa} as well, which reveals that, the corrections to three-phonon-limited $\kappa_L$ by involving four-phonon scattering are more obvious at lower temperatures, and significantly larger in $\beta-$ and $\gamma-$PbP compared to $\alpha-$PbP irrespective of temperatures. The behavior of the temperature-dependence correction $\Delta\kappa_L$, i.e. $\Delta\kappa_L=\kappa_{L,3ph+4ph}-\kappa_{L,3ph}$ is due to the further increase of four-phonon scattering rates compared to three-phonon scattering rates as the temperature increases. 

At low temperature as shown in Figure~\ref{fig:kappa}, the correction $\Delta\kappa_L$ in $\alpha-$PbP is smaller than those in $\beta-$ and $\gamma-$PbP, which is probably due to the relatively smaller four-phonon scattering rates of quasi-acoustic phonon modes compared to the corresponding three-phonon scattering rates in $\alpha-$PbP, since the $\kappa_L$ for all three monolayers are dominantly contributed from quasi-acoustic phonon modes as mentioned above. The relatively smaller correction $\Delta\kappa_L$ in $\alpha-$PbP at high temperatures can be understood in the same way. The calculated $\kappa_L$ at 300 K by considering three- and four-phonon scattering is 0.58 $W/mK$, 0.38 $W/mK$ and 1.04 $W/mK$ for $\alpha$-PbP, $\beta$-PbP and $\gamma$-PbP respectively, which is far below the $\kappa_L$ for other group-\uppercase\expandafter{\romannumeral4} and group-\uppercase\expandafter{\romannumeral5} hexagonal monolayer. e.g. 20-30 $W/mK$ for silicene, 106.6 $W/mK$ for black phosphorene, 5.8 $W/mK$ for Stanene\cite{Issi2014,Nika2009,Lindsay2010,Li2012,Pei2013,Hu2013,Fugallo2014,Lindsay2014,Fugallo2014a,Gu2015,Xie2016,Peng2016}, and about two orders of magnitude smaller than the graphene (2000-5000 $W/mK$). 

Based on the calculated electronic and phonon transport properties, the dimensionless thermoelectric figure of merit $zT$ for $\alpha-$, $\beta-$ and $\gamma-$PbP are calculated and shown in Fig~\ref{te}(g-i). For $\alpha$-PbP, the obtained $zT$ under CRTA reaches 1.85 at E$_f$ = 0.17 eV, which is significantly corrected to XX by considering the \textit{el-ph} interactions in the electronic transport properties. It is noteworthy that, as the extra term of energy dependent relaxation time shows up, the Seebeck coefficient in \textit{el-ph} is little larger than CRTA, which, along with a comparable electronic conductivity below the Fermi level, contributes to a significant $zT$ that reaches 0.90 at E$_f$ = -0.31eV. For $\beta$-PbP, the $zT$ in \textit{el-ph} can hardly catch up with the CRTA results due to an ultra-low electronic conductivity. The highest $zT$ reaches 0.24 at E$_f$ = -0.38eV. For $\gamma$-PbP, our calculation demonstrates a local maximun of 1.25 at E$_f$ = 0.21eV, in consideration of full \textit{el-ph} interaction. The TE merit $zT$ for the three investigated 2D materials is comparable with conventional TE materials like $\mathrm{Bi}_{2} \mathrm{Te}_{3}$ (1.2)\cite{Poudel2008}, $\mathrm{PbTe}$ (0.30)\cite{Zhang2013}, and $\mathrm{SnSe}$ (0.70)\cite{Wang2015a}.

\subsection{Optical properties: excitonic behaviors and optical selection rules}

\begin{figure}[ht!]
\centering
\includegraphics[width=1\linewidth]{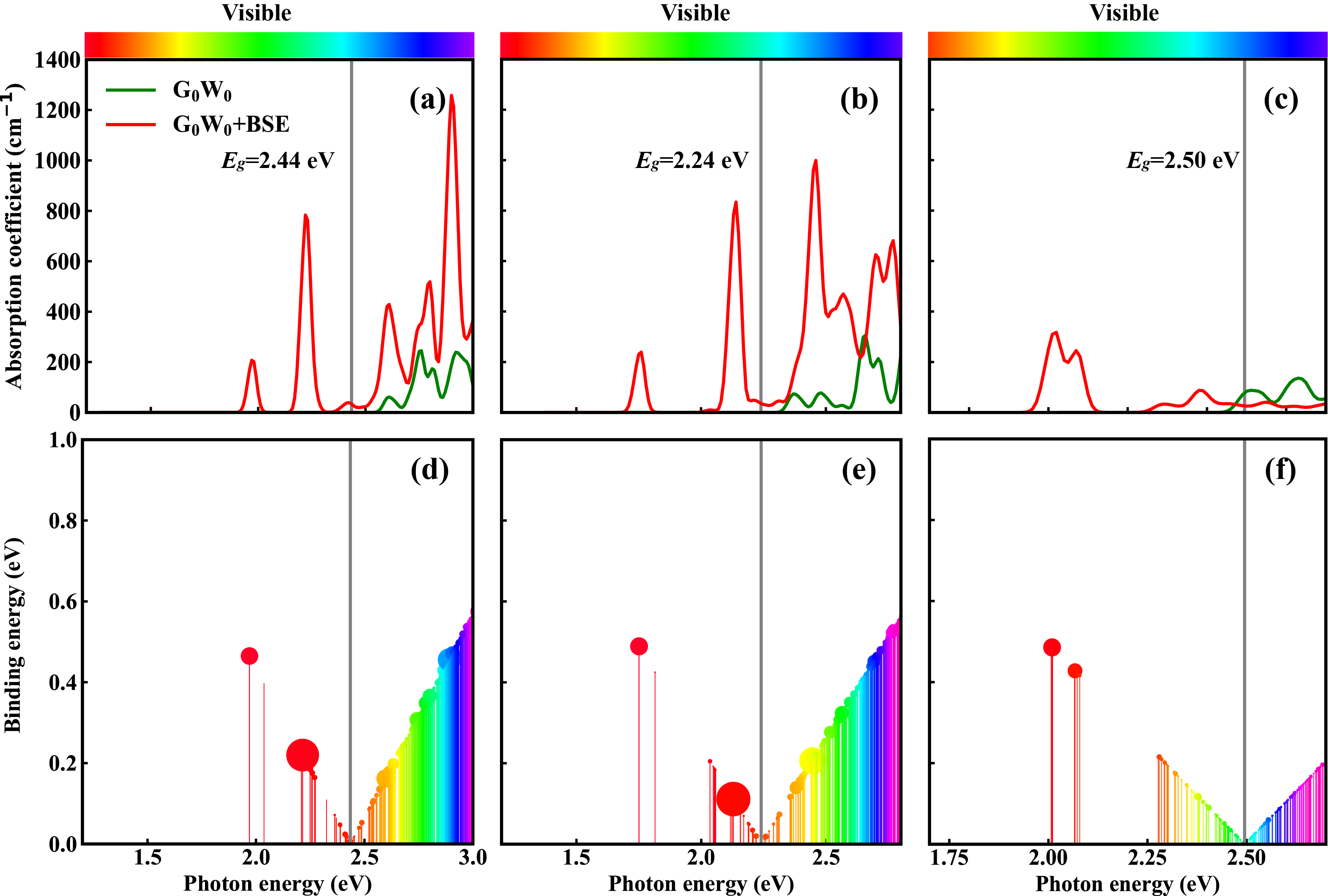}
\caption{(a-c) Optical absorption spectra obtained by G$_\mathrm{0}$W$_\mathrm{0}$+RPA (green lines) and G$_\mathrm{0}$W$_\mathrm{0}$+BSE (red lines) for $\alpha$-, $\beta$- and $\gamma$-PbP, respectively. (d-f) Corresponding excitons binding energies where circle radius imply the oscillator strengths for excitons that contribute dominantly to major optical absorption peaks.}
\label{optical} 
\end{figure}

As a result of the strong dielectric screening effect, electron-hole pairs (excitons) effects contribute dominantly in optical properties in 2D systems.The contribution to optical properties from the $S^{th}$ exciton is indicated by its oscillator strength, which can be written as $f_{S}=\frac{2|\mathbf{e} \cdot\langle 0|\mathbf{v}| S\rangle|^{2}}{\Omega^{S}}$, where $\mathbf{e}$ is photon polarization vector, $\mathbf{v}$ is the velocity operator, $\Omega^{S}$ is the exciton energy for the $S^{th}$ exciton. The excitonic state is written as $|S\rangle=\sum_{\mathbf{k}, v, c} A_{v c \mathbf{k}}^{s}|c \mathbf{k}\rangle \otimes|v \mathbf{k}\rangle$, the summation over several electron-hole pairs. Generally, large oscillator strengths represent large contributions to the optical properties from certain exciton. Fig.~\ref{optical} is the calculated optical absorption coefficient with and without the consideration of excitonic effect for $\alpha$-, $\beta$- and $\gamma$-PbP. The circle radii in (d-f) correspond to the excitonic oscillator strengths. Due to a narrow bandgap, the absorption peaks for the three material locates in visible band.

\section{Conclusion}

\section*{NUMERICAL METHOD AND COMPUTATIONAL DETAILS}

\subsection*{Ground-state calculations}

The $ab$ $initio$ calculations are based on the density funtional theory (DFT) and implemented by the Vienna $ab$ $initio$ simulation package (VASP)\cite{Kresse1996}. The Perdew-Burke-Ernzerhof (PBE) gerneralized gradient approximation (GGA) is used as the exchange-correlation functional and projector-augmented-wave (PAW) pseudopotential is chosen with a cutoff kinetic energy set to 600 eV\cite{Perdew1996}. To reach convergence, a 15$\times$15$\times$1 Monkhorst-Pack\cite{Monkhorst1976} k-mesh is used to sampling the Brillouin zone during the structural optimization and self-consistent calculations for electronic systems. The convergence certerion is -10$^{-4}$ eV/$\mathring{A}$ and $10^{-7}$ eV for Feynman force convergence and electronic energy differences, respectively. 

The $ab$ $initio$ molecular dynamics (AIMD) simulation is used with a 5$\times$5$\times$1 supercell, a time interval of 2 $fs$ and 2000 steps during the simulation. The phonon dispersions with harmonic inter-atomic force constants (IFCs) are calculated via density functional perturbation theory (DFPT) implemented with VASP and Phonopy codes\cite{Togo2008} with a 5$\times$5$\times$1 supercell and 1$\times$1$\times$1 q-mesh\cite{Nose1984,Peng2018}. The Debye temperature can be calculated through 

\subsection*{CALYPSO}
The crystal structures of PbP monolayers are obtained by Calypso (Crystal structure AnaLYsis by Particle Swarm Optimization), a software based on swarm intelligence algorithm, which is able to search stable crystal structures according to their enthalpy\cite{Wang2010,Wang2012}. In the searching, an unit-cell with two Pb and P atoms is considered with two sublayers and a vacuum gap of 30 $\r{A}$. In the first step, 30 structures are generated randomly within the 270 crystal space groups. After that, VASP code is used to optimized the structures until the energy changes within 2$ \times$ 10$^{-3}$ $\mathrm{eV}$. 60$\%$ structures with lower Gibbs free energies are chosen to generate the next generation of structures, according to particle swarm algorithm, while the rest 40$\%$ (12 for 30 structures a generation) is regenerated in crystallographic symmetry scheme. In addition, structure fingerprinting technique of bond characterization matrix is used to avoid equivalent structures. In our case of PbP monolayer, the structures reach a lowest enthalpy within 18 generations. After excluding the unrational structures, we obtained three low-buckled structures with two sublayers, namely $\alpha$-, $\beta$- and $\gamma$-PbP. The enthalpy of formation for the three monolayers is -4.65 eV, -4.63 eV and -4.75 eV, respectively.

\subsection*{Thermal conductivity considering three- and four-phonon interactions}

For materials with a large $a-o$ phonon bandgap, both three- and four-phonon scattering channels should be considered. The three- and four-phonon scattering rates, i.e. $\tau_{3, \lambda}^{-1}$ and $\tau_{4, \lambda}^{-1}$, can be calculated based on the Fermi's golden rule (FGR), written as,

\begin{equation}
\tau_{3, \lambda, \mathrm{RTA}}^{-1}=\sum_{\lambda_{1} \lambda_{2}}\left[\frac{1}{2}\left(1+n_{\lambda_{1}}^{0}+n_{\lambda_{2}}^{0}\right) \mathcal{L}_{-}+\left(n_{\lambda_{1}}^{0}-n_{\lambda_{2}}^{0}\right) \mathcal{L}_{+}\right]
\end{equation}

\begin{equation}
\tau_{4, \lambda, \mathrm{RTA}}^{-1}=\sum_{\lambda_{1} \lambda_{2} \lambda_{3}}\left[\frac{1}{6} \frac{n_{\lambda_{1}}^{0} n_{\lambda_{2}}^{0} n_{\lambda_{3}}^{0}}{n_{\lambda}^{0}} \mathcal{L}_{--}+\frac{1}{2} \frac{\left(1+n_{\lambda_{1}}^{0}\right) n_{\lambda_{2}}^{0} n_{\lambda_{3}}^{0}}{n_{\lambda}^{0}} \mathcal{L}_{+-}+\frac{1}{2} \frac{\left(1+n_{\lambda_{1}}^{0}\right)\left(1+n_{\lambda_{2}}^{0}\right) n_{\lambda_{3}}^{0}}{n_{\lambda}^{0}} \mathcal{L}_{++}\right]
\end{equation}

where $n^{0}=\frac{1}{e^{\hbar\omega_\lambda / k_{B} T}-1}$ is the occupation by Boson-Einstein distribution for phonons with $\omega_\lambda$. The tansition probability matrix $\mathcal{L}$ denotes the transition probability, which is determined by 3$^{rd}$-order and 4$^{th}$-order IFCs\cite{Feng2016}, can be given by the FGR. For three-phonon scatterings, the 3$^{rd}$-order IFCs give the tansition probability matrix $\mathcal{L}^{(3)}$ as,

\begin{equation}
\begin{aligned}
\mathcal{L}_{\pm}^{(3)} 
		=\frac{\pi \hbar}{4 N_{\mathbf{q}}}\left|V_{\pm}^{(3)}\right|^{2} \Delta_{\pm} \frac{\delta\left(\omega_{\lambda} \pm \omega_{\lambda_{1}}-\omega_{\lambda_{2}}\right)}{\omega_{\lambda} \omega_{\lambda_{1}} \omega_{\lambda_{2}}}
\end{aligned}
\end{equation}.

where $N_{\mathbf{q}}$ is the total number of $\mathbf{q}$ points, the Kronecker $\Delta_{\pm}$ denotes the momentum conservation, i.e. $\Delta_{\pm}=\delta(\mathbf{q} \pm \mathbf{q_1}-\mathbf{q_2})$ and $\Delta_{\pm}=\delta(\mathbf{q} \pm \mathbf{q_1}-\mathbf{q_2}+\mathbf{R})$, and $V_{\pm}^{(3)}$ is defined as,

\begin{equation}
V_{\pm}^{(3)}=\sum_{b l_{1} b_{1}, l_{2} b_{2}} \sum_{\alpha \alpha_{1} \alpha_{2}} \Phi_{0 b,l_1 b_1, l_2 b_2}^{\alpha\alpha_1\alpha_2} \frac{e_{\alpha b}^{\lambda} e_{\alpha_{1} b_{1}}^{\pm \lambda_{1}} e_{\alpha_{2} b_{2}}^{-\lambda_{2}}}{\sqrt{\bar{m}_{b} \bar{m}_{b_{1}} \bar{m}_{b_{2}}}} e^{\pm i \mathbf{q}_{1} \mathbf{r}_{l_{1}}} e^{-i \mathbf{q}_{2} \boldsymbol{r}_{l_{2}}}
\end{equation}

where $\Phi_{0 b,l_1 b_1, l_2 b_2}^{\alpha\alpha_1\alpha_2}$ is the 3$^{rd}$ IFCs. For four-phonon scatterings, the 4$^{th}$-order IFCs give the tansition probability matrix $\mathcal{L}^{(4)}$ as,

\begin{equation}
\begin{aligned}
\mathcal{L}_{\pm \pm}^{(4)} 
=\frac{\pi \hbar}{4 N} \frac{\hbar}{2 N_{\mathbf{q}}}\left|V_{\pm \pm}^{(4)}\right|^{2} \Delta_{\pm \pm} \frac{\delta\left(\omega_{\lambda} \pm \omega_{\lambda_{1}} \pm \omega_{\lambda_{2}}-\omega_{\lambda_{3}}\right)}{\omega_{\lambda} \omega_{\lambda_{1}} \omega_{\lambda_{2}} \omega_{\lambda_{3}}}
\end{aligned}
\end{equation}

where $\Delta_{\pm \pm}$ represents $\Delta_{\mathrm{q}+\mathrm{q}_{1}+\mathrm{q}_{2}+\mathrm{q}_{3}, \mathbf{R}}$, indicating the quasi-momentum conservation criterion, and $V_{\pm}^{(4)}$ is defined as,

\begin{equation}
V_{\pm \pm}^{(4)}=\sum_{b, l_{1} b_{1}, l_{2} b_{2}, l_{3} b_{3}} \sum_{\alpha \alpha_{1} \alpha_{2} \alpha_{3}} \Phi_{0 b, l_{1} b_{1}, l_{2} b_{2}, l_{3} b_{3}}^{\alpha \alpha_{1} \alpha_{2} \alpha_{3}} \frac{e_{\alpha b}^{\lambda} e_{\alpha_{1} b_{1}}^{\pm \lambda_{1}} e_{\alpha_{2} b_{2}}^{\pm \lambda_{2}} e_{\alpha_{3} b_{3}}^{-\lambda_{3}}}{\sqrt{\bar{m}_{b} \bar{m}_{b_{1}} \bar{m}_{b_{2}} \bar{m}_{b_{3}}}} e^{\pm i \mathbf{q}_{1} \cdot \mathbf{r}_{l_{1}}} e^{\pm i \mathbf{q}_{2} \cdot \mathbf{r}_{l_{2}}} e^{-i \mathbf{q}_{3} \cdot \mathbf{r}_{l_{3}}}
\end{equation}

where $\Phi_{0 b, l_{1} b_{1}, l_{2} b_{2}, l_{3} b_{3}}^{\alpha \alpha_{1} \alpha_{2} \alpha_{3}}$ is 4$^{th}$-order IFCs. 
For convergence, we take into consideration the 11$^{th}$ and 2$^{nd}$ nearest neighbors for the calculations of 3$^{rd}$- and 4$^{th}$-order IFCs. The lattice thermal conductivities for the three monolayers are calculated using the ShengBTE code\cite{Li2014} with 120$\times$120$\times$1 for 3$^{rd}$-order and 30$\times$30$\times$1 for 4$^{th}$-order interactions. The iterative scheme to solve the Boltzmann-transport equation (iBTE) is used to investigate the three-phonon interactions, while four-phonon interaction, the RTA method (RTA-BTE) is used\cite{Lindsay2013}.

\subsection*{Full electron-phonon couplings}

The mode-resolved scatterings rates for the full \textit{el-ph} couplings by considering the Fan-Migdal (FM) interactions can be written as\cite{Giustino2007},

\begin{equation}
\begin{aligned}
\frac{1}{\tau_{n \mathbf{k}}}=& 2 \operatorname{Im} \sum_{n \mathbf{k}}^{\mathrm{FM}}(\omega)=\frac{2 \pi}{\hbar} \sum_{m \nu} \int \frac{\mathrm{d} q}{\Omega_{\mathrm{BZ}}}\left|g_{m n \nu}(\mathbf{k}, \mathbf{q})\right|^{2} 
		& \times\left[\left(1-f_{m \mathbf{k}+\mathbf{q}}^{0}+n_{\mathbf{q} \nu}\right) \delta\left(\varepsilon_{n \mathbf{k}}-\varepsilon_{m \mathbf{k}+\mathbf{q}}-\hbar \omega_{\mathbf{q} \nu}\right)\right.\\
		&\left.+\left(f_{m \mathbf{k}+\mathbf{q}}^{0}+n_{\mathbf{q} \nu}\right) \delta\left(\varepsilon_{n \mathbf{k}}-\varepsilon_{m \mathbf{k}+\mathbf{q}}+\hbar \omega_{\mathbf{q} \nu}\right)\right]
	\end{aligned}
\end{equation}.

The sum is over all the possible electron initial/final states band index $m/n$ and all the phonon modes index $\mu$ and wavevector $\textbf{q}$. $n_{\mathbf{q} \nu}$ and $f_{m \mathbf{k}+\mathbf{q}}^{0}$ respectively denote the occupation number of phonons and electrons under Bose-Einstein distribution and Fermi-Dirac distribution, respectively. $\omega_{\mathbf{q} \nu}$ is the phonon frequency and $\epsilon_{n \mathbf{k}}$ is the electron eigenvalue with band index $m$ and wavevector $\mathbf{k}$. The \textit{el-ph} matrix element $g_{m n \nu}(\mathbf{k}, \mathbf{q})$ describing the probability of electrons from the initial $\ket{n\mathbf{k}}$ state to the final $\ket{m\mathbf{k+q}}$ state via phonon $\ket{\mathbf{q}v}$ is defined as\cite{Baroni2001},

\begin{equation}
g_{m n \nu}(\mathbf{k}, \mathbf{q})=\left\langle\psi_{\mathrm{m} \mathbf{k}+\mathbf{q}}\left|\Delta_{\mathbf{q} \nu} V^{\mathrm{KS}}\right| \psi_{n \mathbf{k}}\right\rangle
\end{equation}

where $\psi_{n \mathbf{k}}$ and $\psi_{m \mathbf{k}+\mathbf{q}}$ are the initial and final electron states. $\Delta_{\mathbf{q} \nu} V^{\mathrm{KS}}$ is the $\mathbf{q}$ phonon-induced perturbation to the Kohn-Sham (KS) potential. To accomplish an accurate calculation, the Wannierization method is used to interpolate the electronic and phonon band structure in fine $\textbf{k}$- and $\textbf{q}$-mesh of $120 \times 120 \times 1$, implemented by Wannier90 and EPW code\cite{Noffsinger2010,Ponce2016,Ponce2020}.

\subsection*{The DPA method and solution to electronic transport properties}

The DPA method proposed by Bardeen and Shockley is used to calculate the carrier mobilities, in which only the LA phonons in the longwavelength limit are considered\cite{Bardeen1950}. The mobilities for 2D semiconductors can be calculated as\cite{Long2011,Chen2013,Wang2015,Xu2017,Peng2018a},

\begin{equation}
\mu^{2 \mathrm{D}}=\frac{2 e \hbar^{3} C^{2 \mathrm{D}}}{3 k_{\mathrm{B}} T m^{* 2} D_{1}^{2}}
\end{equation}

where $C^{2 D}$ is the 2D elastic modulus, defined by 

\begin{equation}
C^{2 \mathrm{D}}=\frac{\partial^{2} E}{\partial\left(\frac{\Delta l}{l_{0}}\right)^{2}} \times \frac{1}{S_{0}}
\end{equation}

where E is total energy, $\Delta l$ is the spatial change of lattice constant $l_{0}$ under external strains, $S_{0}$ is the are of unit cell. $D_{1}$ is the deformation potential constant for electrons or holes, which describes the strength of the interaction between electrons and LA phonons, given by

\begin{equation}
D_{1}^{\mathrm{e} / \mathrm{h}}=\frac{\Delta E_{\mathrm{CBM} / \mathrm{VBM}}}{\Delta l / l_{0}}
\end{equation}

where $\Delta E_{\mathrm{CBM} / \mathrm{VBM}}$ denotes the energy change of CBM or VBM when applying strain $\Delta l$. $m^{*}$ is the effective mass for carriers, which can be calculated by,

\begin{equation}
	m^{*}=\hbar^{2}\left(\frac{\partial^{2} E(k)}{\partial k^{2}}\right)^{-1}
\end{equation}

To further investigation the electronic transport properties, the semiclassical BTE method based on the rigid-band approximation is used to calculate the temperature- and doping-dependent electrical transport properties, including carrier concentrations $n_{\mathrm{h} / \mathrm{e}}$ for holes/electrons, electronic conductivity $\sigma$, electronic thermal conductivities $\kappa_{\mathrm{e}}$ and Seebeck coefficient $S$, given by\cite{Madsen2006,Yang2008,Hong2016},

\begin{equation}
n_{\mathrm{h}}(T, \mu)=\frac{2}{\Omega} \iint_{\mathrm{BZ}}\left[1-f_{0}(T, \varepsilon, \mu)\right] D(\varepsilon) \mathrm{d} \varepsilon
\label{eq:nh}
\end{equation}

\begin{equation}
n_{\mathrm{e}}(T, \mu)=\frac{2}{\Omega} \iint_{\mathrm{BZ}} f_{0}(T, \varepsilon, \mu) D(\varepsilon) \mathrm{d} \varepsilon
\label{eq:ne}
\end{equation}

\begin{equation}
\sigma_{\alpha \beta}(T, \mu)=\frac{1}{\Omega} \int \bar{\sigma}_{\alpha \beta}(\varepsilon)\left[-\frac{\partial f_{0}(T, \varepsilon, \mu)}{\partial \varepsilon}\right] \mathrm{d} \varepsilon
\label{eq:sigma}
\end{equation}

\begin{equation}
\kappa_{\mathrm{e} \alpha \beta}(T, \mu)=\frac{1}{e^{2} T \Omega} \int \bar{\sigma}_{\alpha \beta}(\varepsilon)(\varepsilon-\mu)^{2}\left[-\frac{\partial f_{0}(T, \varepsilon, \mu)}{\partial \varepsilon}\right] \mathrm{d} \varepsilon
\label{eq:kappae}
\end{equation}

\begin{equation}
S_{\alpha \beta}(T, \mu)=\frac{1}{e T \Omega \sigma_{\alpha \beta}(T, \mu)} \int \bar{\sigma}_{\alpha \beta}(\varepsilon)(\varepsilon-\mu)\left[-\frac{\partial f_{0}(T, \varepsilon, \mu)}{\partial \varepsilon}\right] \mathrm{d} \varepsilon
\label{eq:seebeck}
\end{equation}

Here $\Omega$ is the volume for unit cell, $f_{0}$ denotes the  Fermi-Dirac distribution, $\mu$ is the chemical potential, $D(\varepsilon)$ is the density of states as the functional of energy. $\bar{\sigma}_{\alpha \beta}(\varepsilon)$ is the energy dependent conductivity tensor and can be obtained by $\bar{\sigma}_{\alpha \beta}(\varepsilon)=\frac{1}{N} \sum_{n, \mathbf{k}} \bar{\sigma}_{\alpha \beta}(n, \mathbf{k}) \frac{\delta\left(\varepsilon-\varepsilon_{n, k}\right)}{\mathrm{d} \varepsilon}$, where $N$ is the number of $\textbf{K}$ points and $\bar{\sigma}_{\alpha \beta}(n, \mathbf{k})$ can be calculated through kinetic theory, $i$.$e$. $\bar{\sigma}_{\alpha \beta}(n, \mathbf{k})=$$e^{2} \tau_{n \mathbf{k}} \nu_{\alpha}(n, \mathbf{k}) v_{\beta}(n, \mathbf{k})$. The carrier velocity $v_{\alpha, \beta}(n, \mathbf{k})$ can be calculated by $v_{\mathrm{i}}=\frac{l}{\hbar} \frac{\partial \varepsilon_{n, \mathbf{k}}}{\partial k_{\mathrm{i}}}(i=\alpha, \beta)$.

\begin{acknowledgments}
This work is supported by the National Natural Science Foundation of China under Grant No. 11374063, and Shanghai Municipal Natural Science Foundation under Grant Nos. 19ZR1402900.
\end{acknowledgments}


\end{document}